# MIMOSA: Multi-parametric Imaging using Multiple-echoes with Optimized Simultaneous Acquisition for highly-efficient quantitative MRI

Yuting Chen[1,2,3], Yohan Jun[2,3], Amir Heydari[4], Xingwang Yong[2,3], Jiye Kim[5], Jongho Lee[5], Huafeng Liu[1], Huihui Ye[6], Borjan Gagoski[3,7], Shohei Fujita#[2,3] and Berkin Bilgic#[2,3,8]

1. State Key Laboratory of Extreme Photonics and Instrumentation, College of Optical Science and Engineering, Zhejiang University, Hangzhou, China

2. Athinoula A. Martinos Center for Biomedical Imaging, Massachusetts General Hospital, Boston, Massachusetts, USA

3. Department of Radiology, Harvard Medical School, Boston, Massachusetts, USA

4. Industrial Engineering and Management Systems, Amirkabir University of Technology, Tehran, Iran

5. Department of Electrical and Computer Engineering, Seoul National University, Seoul, Republic of Korea

6. School of Communication Engineering, Hangzhou Dianzi University, Hangzhou, China

7. Fetal-Neonatal Neuroimaging & Developmental Science Center, Boston Children's Hospital, Boston, MA, USA

8. Harvard/MIT Health Sciences and Technology, Massachusetts Institute of Technology, Cambridge, Massachusetts, USA

# Shohei Fujita and Berkin Bilgic contributed equally to this work

Word count:  ~ 5000 words

Corresponding author:

Huafeng Liu

Email: liuhf@zju.edu.cn

Address: State Key Laboratory of Extreme Photonics and Instrumentation, College of Optical Science and Engineering, Zhejiang University, Hangzhou 310027, China.

# Abstract

**Purpose:** To develop a new sequence, MIMOSA, for highly-efficient $T_1$, $T_2$, $T_2^*$, proton density (PD), and source separation quantitative susceptibility mapping (QSM).

**Methods:** MIMOSA was developed based on 3D-quantification using an interleaved Look-Locker acquisition sequence with $T_2$ preparation pulse (3D-QALAS) by combining 3D turbo Fast Low Angle Shot (FLASH) and multi-echo gradient echo acquisition modules with a spiral-like Cartesian trajectory to facilitate highly-efficient acquisition. Simulations were performed to optimize the sequence. Multi-contrast/-slice zero-shot self-supervised learning algorithm was employed for reconstruction. The accuracy of quantitative mapping was assessed by comparing MIMOSA with 3D-QALAS and reference techniques in both ISMRM/NIST phantom and in-vivo experiments. MIMOSA's acceleration capability was assessed at R = 3.3, 6.5, and 11.8 in in-vivo experiments, with repeatability assessed through scan-rescan studies. Beyond the 3T experiments, mesoscale quantitative mapping was performed at 750 μm isotropic resolution at 7T.

**Results:** Simulations demonstrated that MIMOSA achieved improved parameter estimation accuracy compared to 3D-QALAS. Phantom experiments indicated that MIMOSA exhibited better agreement with the reference techniques than 3D-QALAS. In-vivo experiments demonstrated that an acceleration factor of up to R = 11.8-fold can be achieved while preserving parameter estimation accuracy, with intra-class correlation coefficients of 0.998 ($T_1$), 0.973 ($T_2$), 0.947 ($T_2^*$), 0.992 (QSM), 0.987 (paramagnetic susceptibility), and 0.977 (diamagnetic susceptibility) in scan-rescan studies. Whole-brain $T_1$, $T_2$, $T_2^*$, PD, source separation QSM were obtained with 1 mm isotropic resolution in 3 min at 3T and 750 μm isotropic resolution in 13 min at 7T.

**Conclusion:** MIMOSA demonstrated potential for highly-efficient multi-parametric mapping.

**Funding information:**

GE Precision Healthcare; National Institutions of Health, Grand/AwardNumbers: R01 EB028797, P41 EB030006, U01 EB026996, R01 EB032378, UG3 EB034875, R21 AG082377, R01 EB034757, S10 OD036263, U24 NS135561, U24 NS137077, R01 MH132610 and U01 DA055353; National Key Research and Development Program of China, Grant/Award Numbers: 2020AAA0109502; National Natural Science Foundation of China, Grant/Award Numbers: U1809204, 61701436; Zhejiang Provincial Natural Science Foundation of China, Grant/Award Numbers: LY22F010007; Zhejiang University Education Foundation

# 1.  INTRODUCTION

Quantitative magnetic resonance imaging aims to estimate physical parameters, such as relaxation time and magnetic susceptibility, to provide information related to the microstructural environment of tissues.[1–3] It has been shown to offer higher sensitivity in detecting pathophysiological changes,[4] enhanced specificity to tissue composition,[5] and improved capability for tissue characterization[6] compared to the conventional contrast-weighted MRI. For instance, $T_1$ mapping is associated with myelin and macromolecular content, and can highlight early demyelination or tissue edema.[7,8] $T_2$ mapping offers the potential to detect and characterize white matter (WM) in cognitive decline and dementia. $T_2^*$ mapping provides insights into tissue composition and iron/myelin deposition. Quantitative Susceptibility Mapping (QSM) estimates the distribution of tissue magnetic susceptibility and has been used for measuring deoxyhemoglobin in the veins, assessing hemorrhage, and guiding therapy.[9] Further specificity to tissue iron and myelin deposition can be gained using susceptibility source separation modeling, whereby paramagnetic and diamagnetic susceptibility sources in the brain can be disentangled with the aid of a $T_2'$ map.[10]

However, standard parameter mapping techniques estimate a single parameter of interest at a time and thus suffer from long acquisition time, low efficiency, and potential misalignment with other complementary acquisitions, hampering their application in clinical and research studies. As many disease processes exhibit a wide range of pathological changes that affect different aspects of tissues, the absence of certain parameter maps may result in the loss of critical information for diagnosis or treatment monitoring. It may also be difficult to relate the changes in a single parameter to a specific pathological process.[11] In contrast, simultaneous multi-parametric mapping[12] comprises various MRI techniques that use a tailored pulse sequence to estimate a set of parameters with complementary information while eliminating the need for co-registration. By integrating multiple parameters with distinct sensitivities to various aspects of tissue pathology, it can help improve the detection and diagnosis of neurological diseases,[13] facilitate early identification of disease-specific impairment patterns,[12] and provide a more comprehensive assessment of brain tissue.

Several MRI techniques have been proposed to achieve simultaneous multi-parametric mapping.[14–17] 3D MR Fingerprinting (MRF) employs pseudorandomized acquisition parameters to enable whole brain $T_1$ and $T_2$ mapping within a time window feasible for research and clinical studies.[15] 3D-quantification using an interleaved Look-Locker acquisition sequence with $T_2$ preparation pulse (3D-QALAS)[18–22] includes $T_2$- and $T_1$-preparation modules and achieves rapid whole brain $T_1$, $T_2$, and PD mapping by encoding these modules using five 3D turbo Fast Low Angle Shot (FLASH)[23] measurements. STrategically Acquired Gradient Echo (STAGE)[16] uses two fully flow compensated double echo gradient echo acquisitions to enable $T_1$, $T_2^*$, QSM, and PD mapping. 3D Echo Planar Time-resolved Imaging (3D-EPTI)[24] combines an inversion-recovery gradient echo (IR-GE) and a variable-flip-angle gradient-and-spin-echo (VFA-GRASE) sequence with an EPTI readout to enable whole-brain simultaneous $T_1$, $T_2$, $T_2^*$, PD and $B_1^+$ mapping. While it yields data that should lend itself to QSM processing, this capability has not been demonstrated. MR Multitasking[25] and mcLARO[26] allows for simultaneous $T_1$, $T_2$, $T_2^*$, and QSM. MRF has also been extended to provide $T_2^*$ and QSM mapping ability.[27] While these techniques are powerful for multi-parametric estimation, they have not been used in source separation modeling whereby para- and diamagnetic susceptibility sources can be disentangled. Paramagnetic (e.g. iron) and diamagnetic (e.g. myelin) susceptibility sources coexist within individual voxels and collectively contribute to the observed susceptibility contrast. Susceptibility source separation can help investigate neurodegenerative diseases, e.g. by differentiating between multiple sclerosis lesions[28] and tracking the pathological alteration in Alzheimer's disease-driven neurodegeneration.[29] Current susceptibility source separation methods can be categorized into $R_2'$- and $R_2^*$-based approaches. $R_2'$-based susceptibility source separation requires additional $T_2$ mapping acquisition,[10] which increases scan time and necessitates image co-registration. $R_2'$-based methods[10,30,31] attempt to approximate the $R_2'$ map directly from the acquired $R_2^*$ map, though this may comprise accuracy[32]. Given that QSM inversion is inherently ill-posed and challenging, incorporating tissue $T_2$ information could improve source separation robustness and fidelity. Therefore, the development of an integrated acquisition sequence capable of simultaneously providing $T_2$, $T_2^*$, and GRE phase information has the potential to advance susceptibility source separation, enabling more robust and efficient quantification of tissue composition.

Additionally, mesoscale quantitative mapping at submillimeter resolution, often performed at 7T due to the advantage of high signal-to-noise ratio (SNR), allows for the characterization of neuroanatomical structures with high spatial resolution.[33] By providing detailed insight into the composition of brain tissue, it can enhance our understanding of disease mechanisms, particularly in neurodegenerative diseases, allowing for earlier detection and more personalized treatment strategies. However, only a few multi-parametric mapping techniques[17,34,35] were implemented at 7T. Developed using Pulseq, MIMOSA is compatible with a range of scanner vendors and software baselines, including Siemens 7T systems, and readily enables mesoscale quantitative mapping, facilitating advanced neuroimaging studies such as cortical layer analysis and fine-grained susceptibility mapping.

In this study, we developed Multi-parametric Imaging using Multiple-echoes with Optimized Simultaneous Acquisition (MIMOSA) by extending our Pulseq 3D-QALAS implementation[22] to enable whole brain, highly-efficient $T_1$, $T_2$, $T_2^*$, PD, QSM, para- and diamagnetic susceptibility mapping. Simulations, phantom, and in-vivo experiments were conducted to validate the accuracy of our proposed method. The main contributions of this work are:

1. We achieved fast and highly efficient data acquisition by designing the sampling strategy and employing zero-shot self-supervised learning[36,37] for joint multi-contrast/-slice reconstruction. Multi-parametric quantitative imaging was performed at 3T at 1 mm isotropic resolution within 3 minutes, achieving up to R = 11.8-fold acceleration. Additionally, we enabled mesoscale multi-parametric quantitative imaging at 7T with 750 μm isotropic voxel size within 13 minutes.

2. We demonstrated the application of magnetic susceptibility source separation from a single, multi-parametric scan.

3. The accuracy, acceleration capability, and repeatability of the proposed method was validated by comparing MIMOSA with 3D-QALAS and the reference standard methods in both ISMRM/NIST phantom and in-vivo experiments.

# 2. METHODS

## 2.1 MIMOSA sequence and signal model

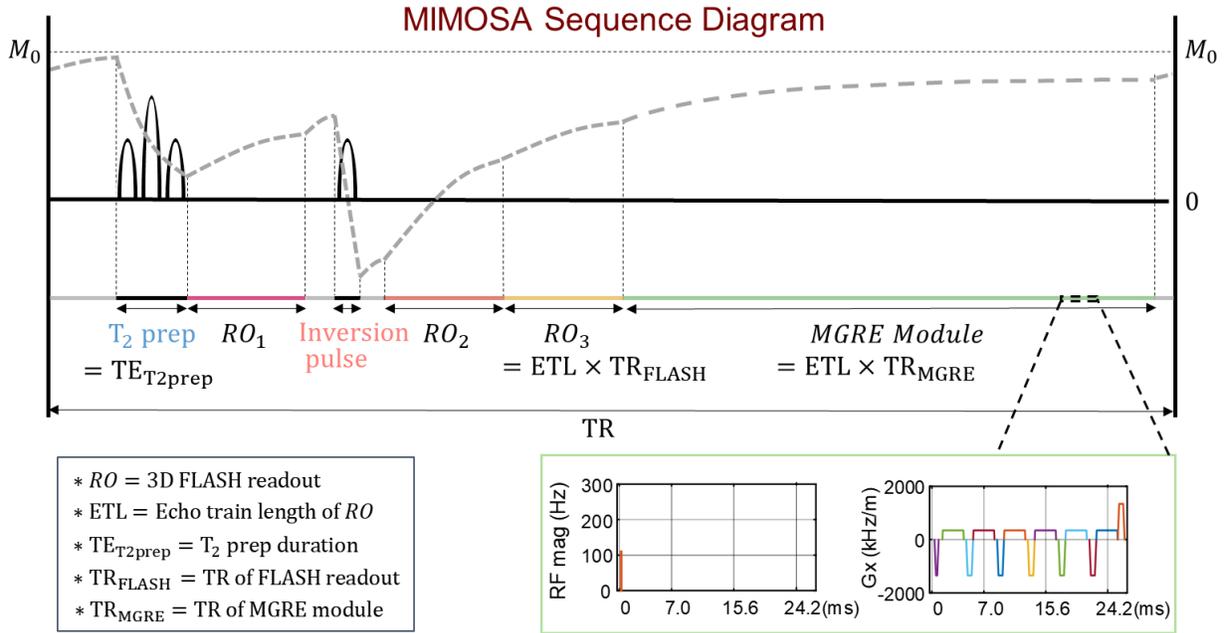

FIGURE 1. Sequence diagram of MIMOSA. MIMOSA consists of a $T_2$ preparation module ($T_2$ prep) followed by one FLASH readout ($RO_1$), and an inversion pulse followed by two FLASH readouts ($RO_{2, 3}$) and one 3D multi-echo gradient echo (MGRE) module including multiple echoes to generate $T_2^*$ and magnetic susceptibility contrast. The dashed box delineates an enlarged view of the MGRE module, detailing the acquisition parameters employed in 3T experiments: TEs = [2.7, 7.0, 11.3, 15.6, 19.9, 24.2] ms and TR = 27.5 ms. The total TR of MIMOSA is 6030 ms.

3D-QALAS[18–22] is a promising multi-parametric mapping technique that was developed using the open-source platform Pulseq[38–40] and demonstrated to have high repeatability and reproducibility in $T_1$, $T_2$, and PD mapping.[22] Using this open-source sequence[22] as a starting point, several augmentations were made to provide $T_2^*$ and susceptibility mapping ability. The newly incorporated multi-echo gradient echo (MGRE) module was optimized to boost the acquisition efficiency and estimation accuracy via simulations (see Section 2.4). Specifically, MIMOSA involves a $T_2$ preparation module followed by one FLASH readout, and an inversion recovery (IR) preparation pulse followed by two FLASH readouts and one MGRE module as shown in Figure 1. One dummy scan was incorporated at the beginning of the sequence to allow the signal to reach steady state. The $T_2$ preparation module consists of one 90° tip-down RF pulse followed by four adiabatic RF refocusing pulses and one 90° tip-up RF pulse. $T_2$ relaxation occurs during this $T_2$-sensitizing period can be described as:

$$M_2 = M_1 \exp\left(-\frac{TE_{T_{2prep}}}{T_2}\right)$$

(1)

where $M_1$ and $M_2$ are the longitudinal magnetizations before and after the $T_2$ preparation module respectively, and $TE_{T2prep}$ is the duration of the $T_2$ preparation block.

Given that previous studies[41,42] have demonstrated superior robustness of Look-Locker based $T_1$ mapping compared to variable flip angle (VFA) methods,[43-46] the same $T_1$-sensitizing scheme as used in 3D-QALAS was adopted. During the $T_1$-sensitizing period, a 180° inversion pulse was applied to encode $T_1$ contrast. Data acquisition was performed using FLASH readout modules consisting of a series of spoiled gradient echo sequences with a flip angle $\alpha$. $T_1$ relaxation occurs between these acquisition modules. The longitudinal magnetization at the $n$+1 th block ($M_{n+1}$) can be calculated based on the magnetization at the $n$th block ($M_n$) before the time interval $\Delta t$ and the initial magnetization ($M_0$) as follows:

$$M_{n+1} = M_n e^{-\Delta t/T_1} + M_0(1 - e^{-\Delta t/T_1})$$

(2)

The steady-state transverse magnetization signal during the MGRE module is given by:

$$M_n(TE_i) = M_0 \cdot \frac{\sin(\alpha)(1 - e^{-TR_{MGRE}/T_1})}{1 - \cos(\alpha)e^{-TR_{MGRE}/T_1}} e^{-TE_i/T_2^*} e^{j(\varphi_0 + 2\pi\Delta f TE_i)}$$

(3)

where $TE_i$ is the echo time, $TR_{MGRE}$ is the repetition time of the MGRE module, $T_2^*$ is the transverse relaxation time, $\varphi_0$ is the phase of magnetization at TE = 0, and $\Delta f$ is the total magnetic field, which includes background field $\Delta f_b$ induced by background susceptibility sources and tissue field $\Delta f_t$ induced by tissue susceptibility sources.[9] $\Delta f_t$ can be obtained by applying phase unwrapping[47,48] and background removal[49,50] on $\Delta f$.

$\Delta f_t$ is modeled as the convolution of field perturbation kernel $D_f$ and the bulk magnetic susceptibility distribution $\chi$. As para- and diamagnetic susceptibility sources typically coexist within individual voxels of brain tissue, $\chi$ can be replaced by the sum of these two sources and $\Delta f_t$ can be expressed as:[10]

$$\Delta f_t = D_f \otimes (\chi_{para} + \chi_{dia})$$

(4)

where $\chi_{para}$ is the paramagnetic susceptibility distribution and $\chi_{dia}$ is the diamagnetic susceptibility distribution.

As $R_2'$ arises from magnetic susceptibility sources, it can be represented as the weighted absolute sum of para- and diamagnetic susceptibility sources as follows:[10]

$$R_2' = D_r \cdot \left(\left|\chi_{para}\right| + \left|\chi_{dia}\right|\right)$$

(5)

where $D_r$ is the relaxometric constant which can be determined by linear regression analysis of deep gray matter regions of interest (ROIs). We adopted the reported value of 114 Hz/ppm for $D_r$ from Kim et al.[31] $R_2'$ also can be derived from $T_2$ and $T_2^*$ using the following relationship:

$$R_2' = \frac{1}{T_2^*} - \frac{1}{T_2}.$$

(6)

Consequently, para- and diamagnetic susceptibility can be derived by solving Equations (4) and (5) given both $R_2'$ and $\Delta f_t$.

A transmit radio frequency magnetic field ($B_1^+$) map was used to correct the nominal flip angle by:

$$\alpha_{cor} = B_1^+ \cdot \alpha$$

(7)

where $\alpha_{cor}$ is the flip angle after correction.

A nonselective adiabatic inversion pulse was applied during the $T_1$-sensitizing period to invert the magnetization. The incomplete inversion of magnetization caused by relaxation during the IR pulse can be captured by inversion efficiency (IE), defined as:

$$M_{n+1} = -\text{IE} \cdot M_n.$$

(8)

MIMOSA was developed on the Pulseq platform using MATLAB (MathWorks, Inc., Natick, MA, USA).

## 2.2 Acquisition and reconstruction

A Variable-Density sampling and Radial view ordering (VDRad)[51] Cartesian k-space sampling trajectory was used for each readout to enable highly efficient data acquisition by cutting the corners of k-space (i.e. elliptical shutter) and using a spiral-like spoke. Each spiral-like spoke contains an echo train length (ETL) of sampling points in the $k_y$-$k_z$ plane and rotates across different repetition times by the golden angle. As a result, complete k-space coverage requires $N_y \times N_z$/ETL TRs, where $N_y$ and $N_z$ represent the phase encoding number in each direction. Distinct sampling patterns were used for different readouts to enable complementary sampling. A center-out k-space sampling order was used in all FLASH readouts, while a reverse out-to-center sampling order was applied in the MGRE module, allowing for longer recovery time and thus higher SNR for $T_2^*$ mapping. Notably, we assumed that image contrast is primarily determined by the echo sampled at the k-space center when generating the dictionary for parameter estimation (see Discussion). This corresponds to the first echo of the echo train in FLASH readouts and the last echo of echo train in the MGRE module.

Multi-contrast/-slice zero-shot self-supervised-learning (MZS-SSL)[30] was used to jointly reconstruct all whole-brain contrasts of MIMOSA. For a fair comparison, the volumes acquired by 3D-QALAS were also reconstructed using MZS-SSL. In accordance with the masking strategy of ZS-SSL, the undersampling mask was partitioned into three disjoint subsets using a randomly uniform distribution for training, validation, and loss calculation (Figure S1 (A)). A set of slice-specific training and loss mask pairs were generated based on slice numbers to promote data incoherence. During training, a single random slice was selected per epoch (Figure S1 (B)), with its multi-contrast data concatenated along the channel dimension and fed into the network. Subsequently, one of the pre-generated training masks was applied to enforce data consistency constraints within the unrolled network which adopts the same network architecture as Zero-DeepSub[21]. The paired loss mask was used to compute the training loss and update network parameters. Validation used only the middle slice with a validation mask to determine when to terminate the iterations (Figure S1 (C)). During inference, the trained model reconstructed whole-brain volumes slice by slice (Figure S1 (D)). The network was trained on an NVIDIA A100 GPU (NVIDIA, Inc., Santa Clara, CA, USA).

## 2.3 Parameter estimation

A dictionary was generated based on the MIMOSA signal model with the following parameter ranges: $T_1$ = [5:10:3000, 3100:50:5000] ms, $T_2$ = [1:2:350, 370:20:1000, 1100:100:3000] ms, and $T_2^*$ = [1:1:100, 105:5:200 210:50:500] ms. To avoid increasing the dictionary size, a look up table about IE was pre-computed based on Bloch simulation which characterizes the magnetization response of the adiabatic IR pulse using the same parameter ranges. $B_1^+$ inhomogeneity was considered by acquiring a $B_1^+$ map using a product turbo-FLASH sequence. The $B_1^+$ values were then quantized within the range of 0.65 to 1.35 and sorted into 50 discrete bins.

After MZS-SSL reconstruction, the magnitude images obtained from the MGRE module were used to estimate $T_2^*$ maps by variable projection (VAPRO),[52] which solves the non-linear fitting problem through signal amplitude projection. The corresponding phase images were used to calculate $\Delta f_t$. Then the estimated $T_2^*$ map served as prior to jointly estimate $T_1$, $T_2$, and PD maps using all acquired contrasts by dictionary matching. N4 bias field correction[53] was applied to correct $B_1^-$ non-uniformity in the PD map. Leveraging the $T_2$ and $T_2^*$ maps, $\chi$-sepnet[31] was employed to calculate total QSM from $\Delta f_t$ and separate the para- and dia-magnetic susceptibility sources.

## 2.4 Simulations

Simulations were conducted to optimize the sequence by evaluating the impact of the number of FLASH readouts and MGRE module positioning. Since 3D-QALAS contains five FLASH readouts (Figure S3 (A)), we started with a setup with four FLASH readouts and one MGRE module, where the acquisition of the first echo in the MGRE module effectively replaces the fifth FLASH readout of 3D-QALAS. To maintain the SNR for $T_2^*$ mapping, the MGRE module was placed after the third FLASH readout or later, ensuring sufficient signal recovery time. Accordingly, the simulations of MIMOSA were performed with the following configurations:

    1) four FLASH readouts followed by the MGRE module (Figure S3 (B));

2) four FLASH readouts with the MGRE module inserted before the last FLASH readout (Figure S3 (C));

3) three FLASH readouts followed by the MGRE module (Figure S3 (D)).

To investigate whether the delays in the original 3D-QALAS between the FLASH readouts were necessary for MIMOSA, the same configurations without delays were also performed (Figure S3, (E-G)).

To generate the reference $T_1$, $T_2$, and $T_2^*$ maps for simulations, typical $T_1$, $T_2$, and $T_2^*$ values of white matter (WM, $T_1/T_2/T_2^*$ = 850/69/40 ms), gray matter (GM, $T_1/T_2/T_2^*$ = 1300/73/45 ms), and cerebrospinal fluid (CSF, $T_1/T_2/T_2^*$ = 4160/1000/200 ms) at 3T were assigned to the corresponding tissues of the BrainWeb digital brain phantom.[54] In the PD map, the relative concentrations of these tissues were set to 0.7, 0.8, and 1, respectively. $B_1^+$ inhomogeneity was considered in the simulations by using the $B_1^+$ map from MR-zero simulation[55]. IE map was generated based on the pre-computed IE look up table. All these maps were shown in Figure S2. Based on the signal model of MIMOSA, the simulated images were generated by Bloch simulation using following parameters: ETL = 127, TR of FLASH readout = 5.8 ms, TE = 2.29 ms, flip angle = 4°, TR of MGRE module = 27.5 ms, TEs of MGRE module = [2.7, 7.0, 11.3, 15.6, 19.9, 24.2] ms. Each FLASH readout has a duration of 736.6 ms, whereas the MGRE module lasted 3492.5 ms. 3D-QALAS simulation[56] was performed using the same parameters with the TR of 4500 ms for benchmarking. Gaussian noise was added to achieve an SNR of 40 dB. All quantitative maps of 3D-QALAS and MIMOSA were estimated by dictionary matching. The normalized root mean squared error (NRMSE) was utilized to quantify the parameter estimation accuracy, defined as:

$$\text{NRMSE} = \frac{\lVert \text{est} - \text{ref} \rVert_2}{\lVert \text{ref} \rVert_2}$$

(9)

where est is the estimated maps and ref is the reference maps.

Furthermore, to evaluate the QSM and susceptibility source separation performance of MIMOSA, numerical phantom was generated using the reference parameter maps obtained from conventional 3D-MGRE and 3D-QALAS methods (first column, Figure S5). Specifically, the reference $T_1$, $T_2$, and PD maps were acquired by 3D-QALAS with the following protocol: FOV = 256×240×224 mm³, resolution = 1×1×1 mm³, ETL = 127, TR of FLASH readout = 5.8 ms, TE = 2.29 ms, flip angle = 4°, acceleration rate R = 7.33, TR = 4500 ms, number of TRs = 58, and total acquisition time (TA) = 5 min 50 sec. FSL FAST[57] was applied to the $T_1$ map to segment WM, GM, and CSF for subsequent quantitative comparisons. The reference $T_2^*$ map was obtained by VAPRO[52] using 3D-MGRE with the following protocol: FOV = 256×240×224 mm³, resolution = 1×1×1 mm³, TEs = [2.7, 7.0, 11.3, 15.6, 19.9, 24.2] ms, flip angle = 15°, acceleration rate R = 6, TR = 28 ms, TA = 3 min 42 sec. The 3D-MGRE phase data were unwrapped,[47] and the background field was removed[58] to obtain the tissue field map. The $T_2$ map from 3D-QALAS was registered to 3D-MGRE using FSL FLIRT[59–61] for estimation of QSM and separation of susceptibility sources by χ-sepnet[31]. Numerical phantom simulations were performed using these parameter maps based on Bloch simulation. The 3D-MGRE simulation was performed using the same TE and TR with the MGRE module of MIMOSA (TEs = [2.7, 7.0, 11.3, 15.6, 19.9, 24.2] ms and TR = 27.5 ms), while 3D-QALAS simulation followed the previous implementation. Complex Gaussian noise was added to achieve an SNR of 40 dB. NRMSE was utilized to quantify the parameter estimation accuracy of $T_1$, $T_2$, $T_2^*$, and PD mapping, while mean absolute error (MAE) was used for QSM and susceptibility source separation evaluation due to their near-zero value ranges.

## 2.5 Phantom experiments

To evaluate the accuracy of $T_1$, $T_2$, and $T_2^*$ mapping with MIMOSA, phantom experiments were performed using an ISMRM/NIST phantom (Premium System Phantom, Caliber MRI, Inc., Boulder, CO, USA) on a 3T MAGNETOM Prisma scanner (Siemens Healthcare, Erlangen, Germany) with a 20-channel head receiver coil. The acquisition parameters of MIMOSA were: FOV = 240×224×192 mm$^3$, resolution = 1×1×4 mm$^3$, BW = 347 Hz/pixel, ETL = 127, TR of FLASH readout = 5.8 ms, TE = 2.29 ms, TR of MGRE module = 27.5 ms, TEs of MGRE module = [2.7, 7.0, 11.3, 15.6, 19.9, 24.2] ms, flip angle = 4°, acceleration rate R = 1.3, total TR = 6030 ms, number of TRs = 86, and TA = 8 min 42 sec. $B_1^+$ maps were acquired using a Siemens product turbo-FLASH[62] sequence and interpolated to match the matrix size. 3D-QALAS with the TR of 4500 ms was acquired using the same protocols and TA = 6 min 27 sec.

For quantitative comparison, the reference $T_1$ and $T_2$ maps were obtained by the product inversion-recovery fast-spin-echo (IR-FSE) and single-echo fast-spin-echo (SE-FSE) sequences, respectively. A $T_2^*$ map was acquired by 3D-MGRE sequence developed using Pulseq. The protocols can be found in Table S2.

## 2.6 In-vivo experiments

To investigate the acceleration capability of MIMOSA, one healthy volunteer (male; age, 26 y) was scanned under three different acceleration rates (R = 3.3, 6.5, and 11.8 for 104, 53, and 29 TRs, respectively). The acquisition parameters of MIMOSA were the same with the phantom experiments except for the resolution (1 mm isotropic). 3D-QALAS was also performed using the same parameters with the TR of 4500 ms as a comparison. Both MIMOSA and 3D-QALAS were reconstructed using the MZS-SSL. The reference $T_1$, $T_2$, and $T_2^*$ maps were obtained using the IR-FSE, SE-FSE, and 3D-MGRE, respectively, following the protocols listed in Table S3. To ensure a fair comparison, matched scan time acquisitions (2 min 55 sec) were also performed for both MIMOSA (R = 11.8 and number of TRs = 29) and 3D-QALAS (R = 8.73 and number of TRs = 39). For quantitative comparison, 4 ROIs (putamen, thalamus, anterior white matter, and posterior white matter) with the size of 5×5 mm$^2$ were extracted from the locations indicated in Figure 5 (D) and Figure S9 (B).

To evaluate the performance of MIMOSA in susceptibility mapping, the reference QSM and susceptibility separation maps were obtained using 3D-MGRE with $T_2$ map from co-registered[57] 3D-QALAS (R = 3.3). Five ROIs (caudate nucleus, globus pallidus, WM, internal capsule, and corpus callosum) were carefully placed by a neuroradiologist and the locations were indicated in Figure S9 (A). The size of each ROI was 5×5 mm$^2$, except for the internal capsule, which was 3×3 mm$^2$ due to its narrower anatomical structure.

To evaluate the repeatability of MIMOSA, a scan-rescan assessment was conducted. 3D-QALAS was performed using the same protocol as a reference. Five healthy volunteers (two male and three female; mean age, 26 y) were scanned twice at R = 6.5 and 11.8. For each subject, the protocol included sequential acquisitions: 3D-QALAS followed by MIMOSA at R = 6.5, and then 3D-QALAS followed by MIMOSA at R = 11.8. A $B_1^+$ map was acquired after these scans. Subjects were then asked to sit up and repositioned for rescan. $B_0$ shimming was performed at the beginning of both the initial and repeat scans to ensure field homogeneity.

All maps from the two scans were registered using FSL FLIRT.[59–61] Seven ROIs (frontal WM, occipital WM, corpus callosum, caudate, thalamus, putamen, and brain stem) with the size of 5×5×5 mm$^3$ were placed by an experienced neuroradiologist. Coefficient of variation (CV) was used to assess the scan-rescan repeatability of $T_1$, and $T_2$, and $T_2^*$ mapping. CV is defined as the standard deviation of the mean parameter values across all ROIs of two subjects between scan and rescan, divided by the corresponding mean value, as expressed in the following equation:

$$\text{CV(MIMOSA/3D-QALAS, R, parameter)} = \frac{\sigma_{\text{scan-rescan}}}{\mu_{\text{scan-rescan}}}.$$  (10)

Intra-class correlation coefficient (ICC) with a two-way mixed-effects model was calculated for the repeatability CV of all ROIs from five subjects.

All in-vivo experiments at 3T were conducted on a Prisma scanner (Siemens Healthcare, Erlangen, Germany) with a 64-channel head receiver coil. Coil compression[63] was used to compress the data to 32-channels.

Further, mesoscale quantitative mapping was performed at 7T. Given the shortened $T_2^*$ values at higher magnetic field strength, we optimized the MGRE module by reducing the number of echoes from six to four. One subject (male; age, 24 y) was scanned with the following parameters: FOV = 240×232×191 mm$^3$, matrix size = 320×310×255, resolution = 750×748×749 μm$^3$, ETL = 115, TR of FLASH readout = 6.3 ms, TE = 2.9 ms, TR of MGRE module = 27.5 ms, TEs of MGRE = [3, 10, 17, 24] ms, flip angle = 4°, TR = 5713 ms, R = 4, number of TRs = 134, and TA = 12 min 57 s. The data was reconstructed using MZS-SSL. The experiment was performed on a 7T Terra scanner (Siemens Healthcare, Erlangen, Germany) with a 64-channels head receiver coil and the data was compressed to 32-channels.

All in-vivo experiments were performed with the approval of the Institutional Review Board.

# 3. RESULTS

## 3.1 Simulations

Table S1 shows the estimation NRMSE for $T_1$, $T_2$, PD, IE, and $T_2^*$ mapping of 3D-QALAS and MIMOSA under six different sequence configurations across three tissue types (WM, GM, and CSF), with the lowest NRMSE for each parameter highlighted in bold. The configuration 1.1 (four FLASH readouts followed by the MGRE module with delays during readouts) yields the lowest NRMSEs for most cases due to the acquisition of one additional volume and longer inversion delay times, providing better conditioning for parameter estimation. The comparison of MIMOSA with and without delays between FLASH readouts was also presented in Table S1 (configurations 1.1 vs. 1.2, 2.1 vs. 2.2, and 3.1 vs. 3.2). Although slight reduction in parameter estimation accuracy for tissues with long $T_1$ relaxation times (particularly CSF) was observed in the configurations without delay, the configuration 3.2 (three FLASH readouts followed by the MGRE module without delays during readouts) achieves lower NRMSEs across all parameters and tissue types compared to 3D-QALAS. It maintained estimation accuracy comparable to the five-readout setup while preserving robustness to $B_1^+$ inhomogeneity and reducing TR by approximately 12% (6.74 vs. 6.03). Therefore, the configuration 3.2 was selected as the final setup for MIMOSA and used in all subsequent experiments.

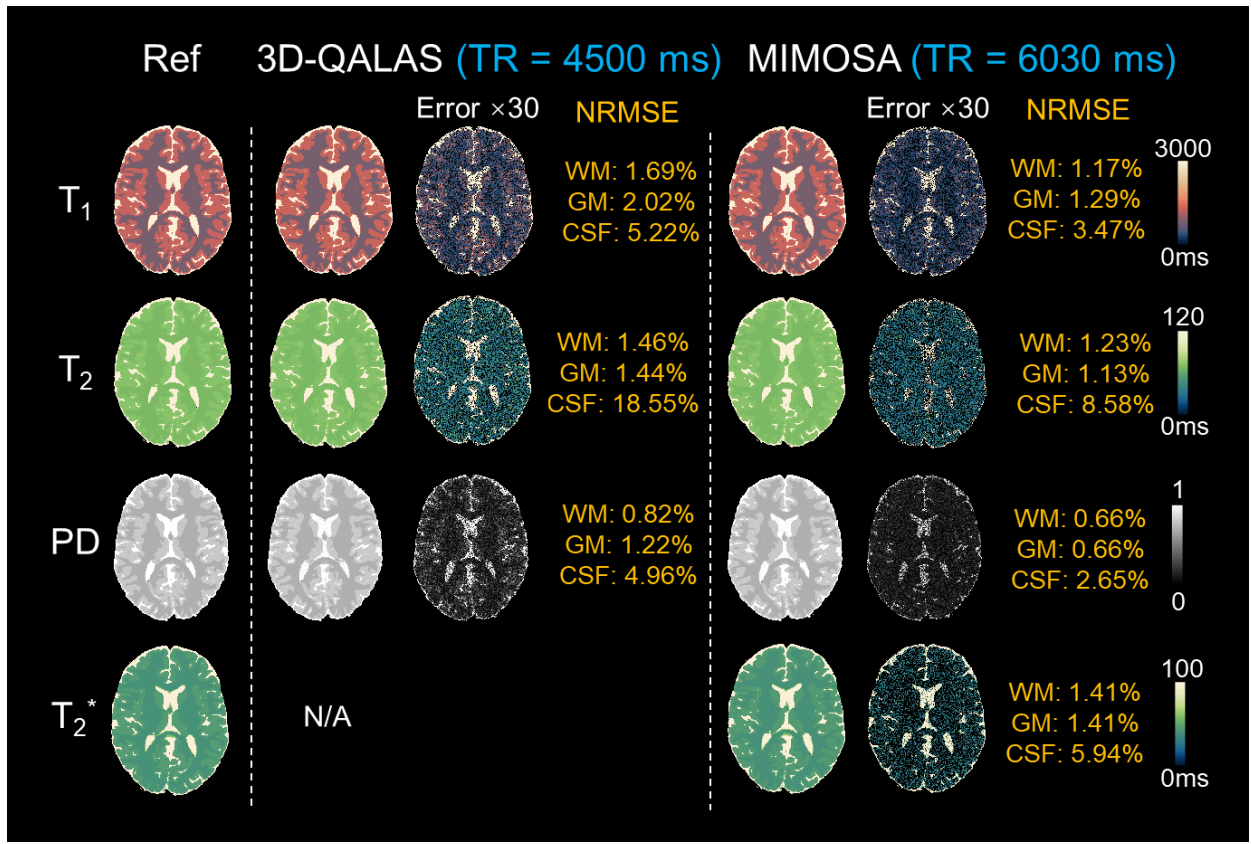

FIGURE 2. Simulation results. The reference $T_1$, $T_2$, PD, and $T_2^*$ maps (first column) were generated using the digital brain phantom from BrainWeb by assigning the typical $T_1$, $T_2$, and $T_2^*$ values of three brain tissues (WM: $T_1/T_2/T_2^* =$

850/69/40 ms, GM: $T_1/T_2/T_2^* = 1300/73/45$ ms, CSF: $T_1/T_2/T_2^* = 4160/1000/200$ ms). For PD map, the relative concentrations of hydrogen protons were used and set to 0.7, 0.8, and 1 respectively. Then the reference maps were used to generate the simulated images of 3D-QALAS and MIMOSA. All quantitative parameter maps (second and fourth column) were estimated by dictionary matching. The absolute error maps (magnified by a factor of × 30) were calculated by comparing the estimated parameter maps with the reference maps and the corresponding NRMSEs of parameter estimation were also provided.

Figure 2 displays the parameter maps and corresponding absolute error maps of 3D-QALAS and MIMOSA using the optimized setup, estimated from the synthesized multi-contrast images (Figure S4). Compared to 3D-QALAS, the $T_1$, $T_2$, and PD maps using MIMOSA show smaller NRMSEs in all three tissues. Additionally, MIMOSA provides an additional $T_2^*$ mapping capability (last row of Figure 2), with the $T_2^*$ estimation NRMSE (WM:1.41%, GM:1.41%, and CSF:5.94%) remaining low. Numerical phantom simulation results (Figure S5) demonstrate that MIMOSA achieves comparable accuracy to 3D-MGRE and 3D-QALAS across all seven parameter maps, with NRMSE less than 6% for relaxation parameters and MAE less than 0.2 ppb for susceptibility parameters.

## 3.2 Phantom validation

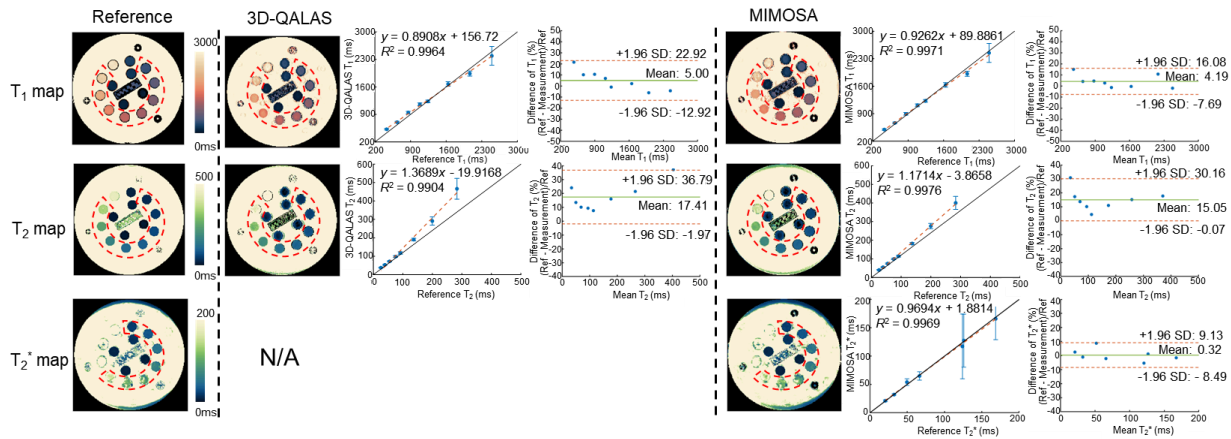

FIGURE 3. Regression and Bland-Altman plots of $T_1$, $T_2$, and $T_2^*$ values in NIST experiments by comparing MIMOSA and 3D-QALAS with reference methods. The reference methods including IR-FSE, SE-FSE, and MGRE were performed and the corresponding $T_1$, $T_2$, and $T_2^*$ mapping results are shown in the first column. For quantitative comparison, the spheres of interest (within the red dotted box) were extracted. In the regression plot, ROI measurement error bars (solid blue lines), a fitting line (red dotted line), and coefficient of determination $R^2$ are provided. In Bland-Altman analysis, the limits of agreement lines are set at ±1.96 standard deviations (SD) from the mean difference of parameter maps, indicating the range within which 95% of the differences between the quantitative maps obtained from MIMOSA or 3D-QALAS and those from the reference methods falls.

Figure 3 compares the $T_1$, $T_2$, and $T_2^*$ maps obtained using proposed MIMOSA on an ISMRM/NIST phantom with those derived from reference methods (IR-FSE, SE-FSE, and MGRE), as well as the $T_1$ and $T_2$ maps obtained using 3D-QALAS. Spherical masks were applied to isolate the spheres of interest containing values within the typical quantification range for brain tissue,[64,65] as indicated by red dotted boxes in Figure 3.

Regression plots are used to assess the linearity of the relationship between the $T_1$, $T_2$, and $T_2^*$ values from MIMOSA and the reference methods (third column in Figure 3), as well as between 3D-QALAS and the same reference acquisitions (sixth column in Figure 3). The regression slopes (0.9262 for $T_1$ and 1.1714 for $T_2$) of MIMOSA demonstrate better linearity compared to those (0.8908 for $T_1$ and 1.3689 for $T_2$) of 3D-QALAS. The coefficient of determination ($R^2$) for $T_1$ is slightly higher in MIMOSA (0.9971) compared to 3D-QALAS (0.9964). MIMOSA exhibits a higher $R^2$ for $T_2$ (0.9976 vs. 0.9904). For $T_2^*$ values, MIMOSA demonstrates a close alignment with the reference MGRE, showing a regression slope of 0.9694 and $R^2$ of 0.9969. Additionally, separate Bland-Altman plots were drawn to assess the agreement between MIMOSA and the reference methods, and between 3D-QALAS and the reference methods. The difference values were calculated as percentages. Specifically, MIMOSA shows the mean biases of 4.19% for $T_1$ mapping and 15.05% for $T_2$ mapping, while higher mean bias was observed in 3D-QALAS with the values of 5.00% for $T_1$ mapping and 17.41% for $T_2$ mapping. In the estimation of $T_2^*$, MIMOSA shows a mean bias of 0.32%, indicating high accuracy.

## 3.3 In-vivo evaluation

### 3.3.1 Acceleration capability validation

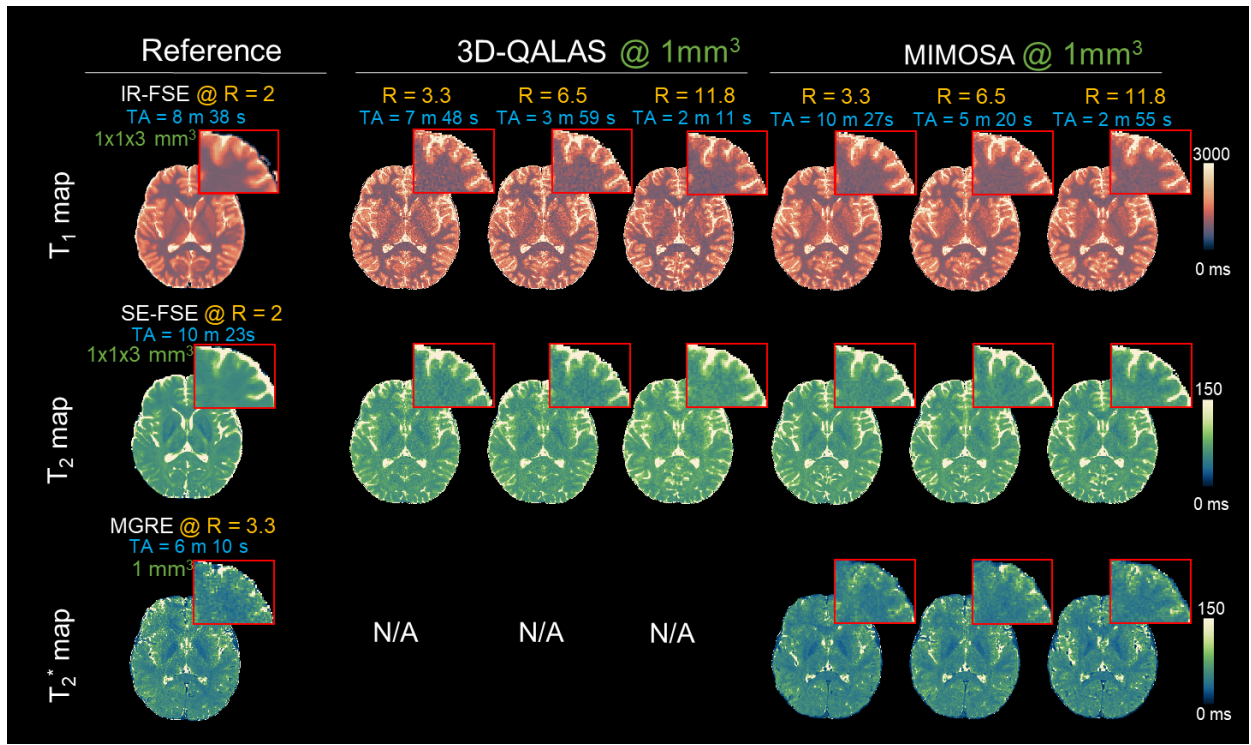

FIGURE 4. The comparison of $T_1$ and $T_2$ maps between MIMOSA and 3D-QALAS at three acceleration rates (R = 3.3, 6.5, 11.8) with 1 mm³ isotropic resolution, using IR-FSE (R = 2), SE-FSE (R = 2), and MGRE (R = 3.3) as reference methods for $T_1$, $T_2$, and $T_2^*$ mapping, respectively. The resolution of IR-FSE and SE-FSE is 1×1×3 mm³ and MGRE is 1 mm³. The corresponding total acquisition times (TA) were provided.

Figure 4 compares the quantitative mapping of 3D-QALAS and MIMOSA with 1 mm isotropic resolution at three different acceleration factors (R = 3.3, 6.5, and 11.8). Both 3D-QALAS and MIMOSA were reconstructed using MZS-SSL. The reconstructed images and corresponding sampling masks of 3D-QALAS and MIMOSA were shown in Figure S7 and S8. The reference $T_1$, $T_2$, and $T_2^*$ maps were presented in the first column of Figure 4. Benefiting from acquisition of additional multi-echo images, MIMOSA qualitatively outperforms 3D-QALAS in image reconstruction quality for both $T_1$ and $T_2$ maps.

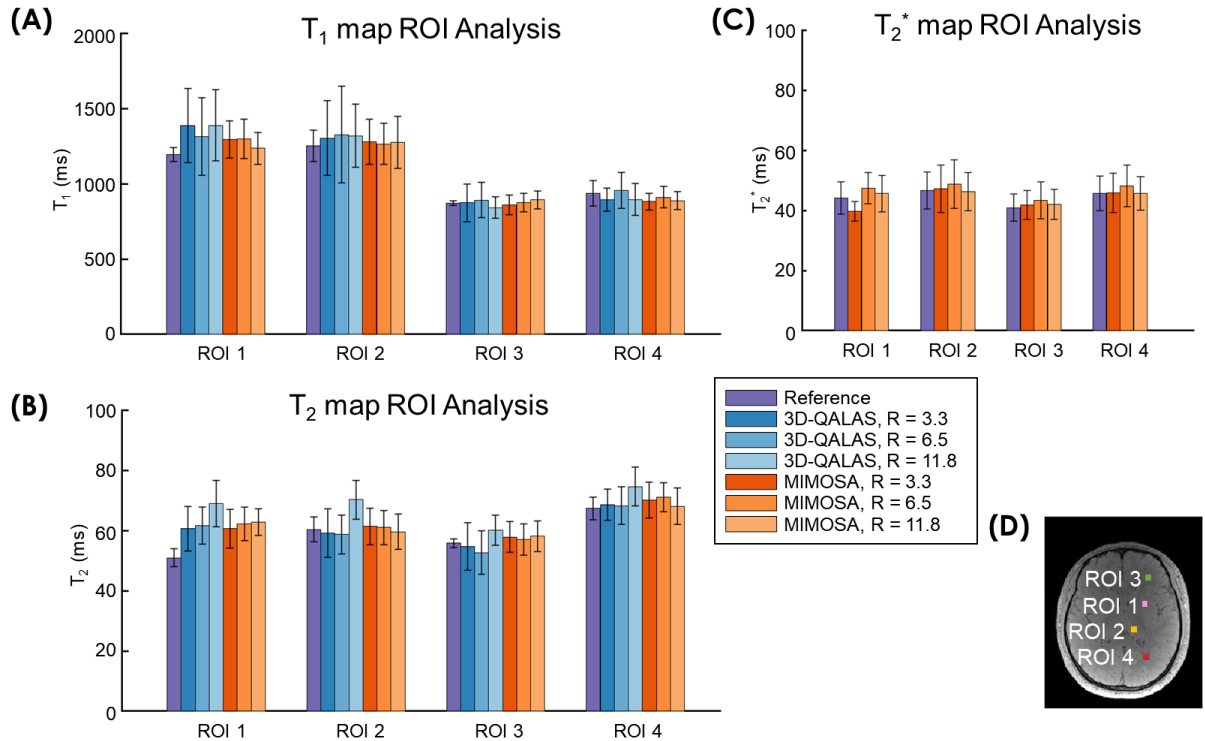

FIGURE 5. ROI analysis for (A) $T_1$ and (B) $T_2$ mapping by comparing the reference methods (IR-FSE and SE-FSE) with 3D-QALAS and MIMOSA under three different acceleration rates (R = 3.3, 6.5, and 11.8). The ROI analysis for $T_2^*$ mapping (C) compares the reference MGRE with MIMOSA under R = 3.3, 6.5, and 11.8. (D) Locations of the placed ROIs.

Figure 5 presents the ROI analysis of the quantitative maps in Figure 4. Compared to 3D-QALAS, MIMOSA demonstrates better agreement with the reference methods for both $T_1$ and $T_2$ mapping, with reduced standard deviations. For $T_2^*$ mapping, MIMOSA shows comparable values and standard deviations to the reference method. $T_1$, $T_2$, and $T_2^*$ mapping at different acceleration rates using MIMOSA exhibits similar values across all four ROIs, demonstrating MIMOSA's ability to retain accuracy despite the fast acquisition. Figure S10 compares $T_1$ and $T_2$ mapping between MIMOSA and 3D-QALAS at matched scan times, demonstrating that MIMOSA preserves comparable parameter estimation performance to 3D-QALAS while providing additional $T_2^*$ and susceptibility mapping capability. Figure 6 presents a comparison of QSM and susceptibility maps between MIMOSA and reference methods, illustrating that MIMOSA provides comparable accuracy in susceptibility mapping and source separation even at high acceleration rate (R = 11.8). Figure 7 shows whole-brain $T_1$, $T_2$, $T_2^*$, PD, QSM, para- and diamagnetic susceptibility maps using MIMOSA at R = 11.8.

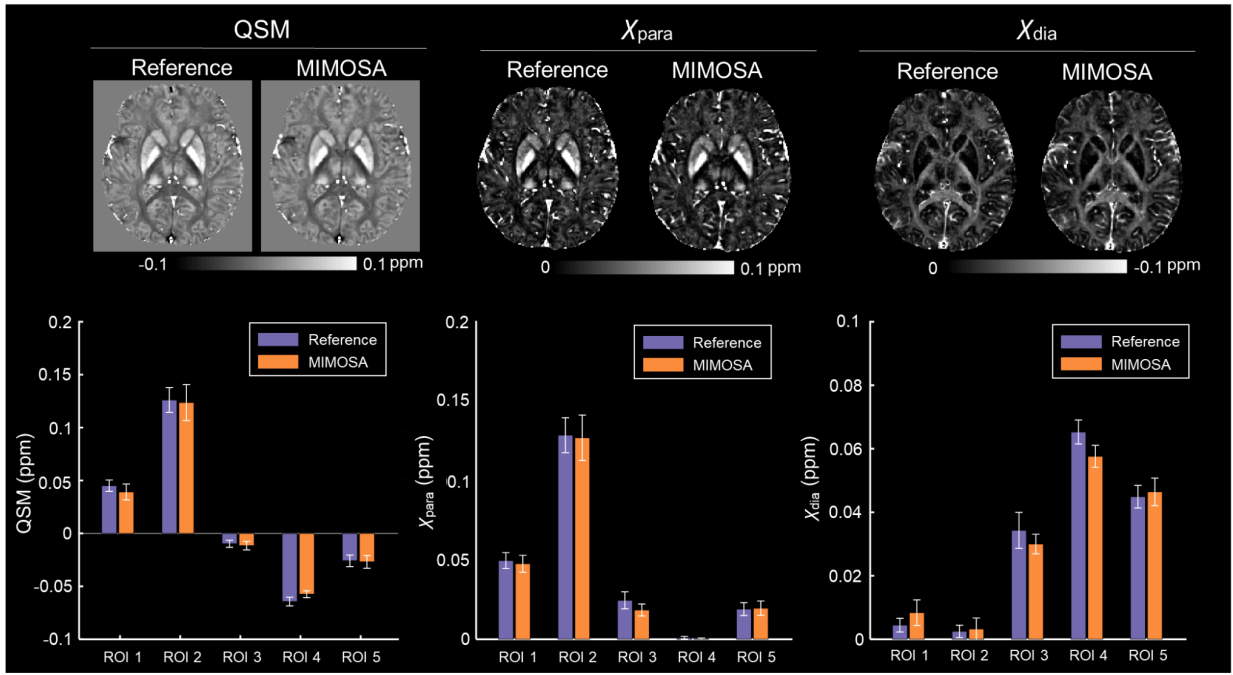

FIGURE 6. Comparison of QSM, paramagnetic, and diamagnetic susceptibility maps between the reference method (3D-MGRE + $T_2$ map of 3D-QALAS, R = 3.3) and MIMOSA (R = 11.8). Five ROIs (Figure S9 (A)) were selected for quantitative analysis.

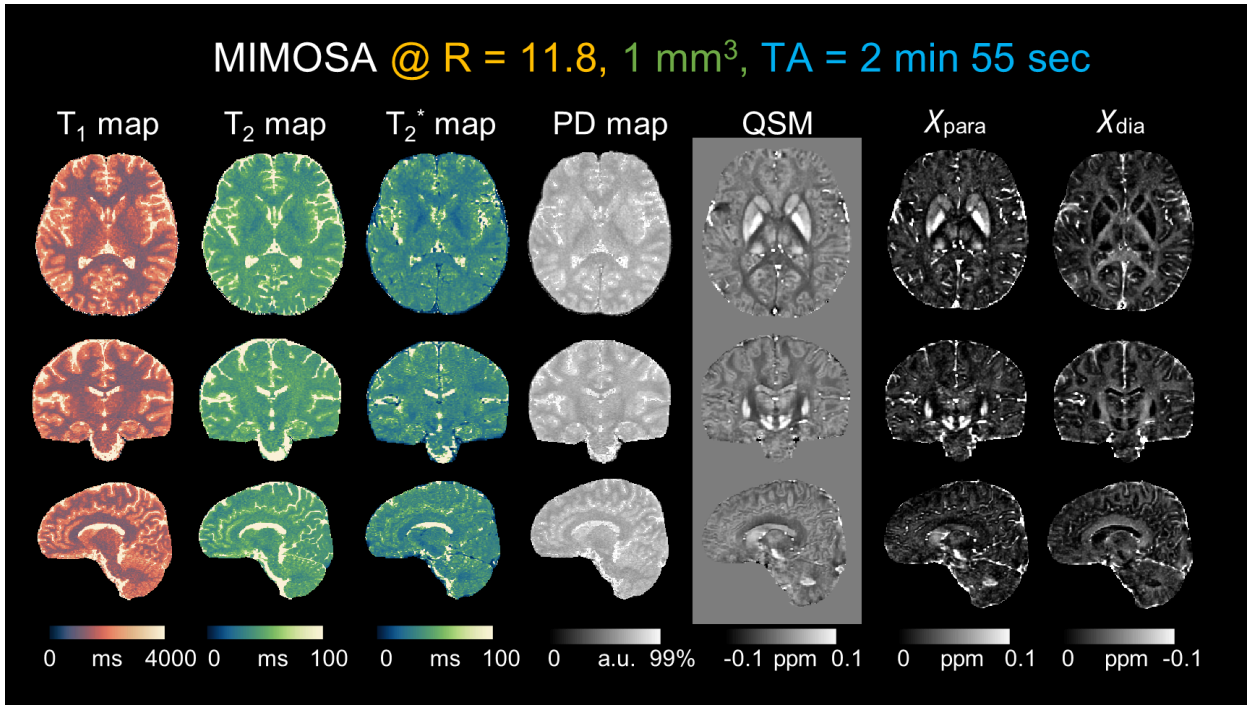

FIGURE 7. Simultaneous whole-brain $T_1$, $T_2$, $T_2^*$, PD, QSM, para- and diamagnetic susceptibility mapping at 1-mm isotropic resolution in 2 minutes 55 seconds with an acceleration rate of 11.8 using MIMOSA.

## 3.3.2 Scan-rescan repeatability evaluation

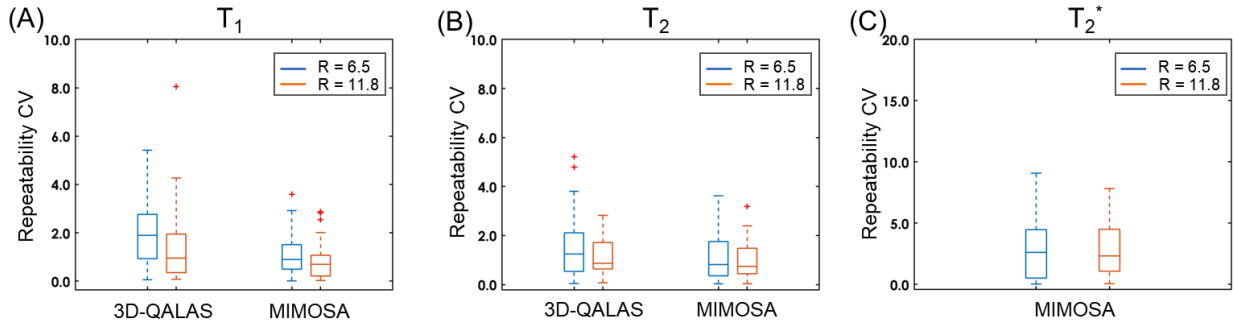

FIGURE 8. Scan-rescan repeatability CV of $T_1$, $T_2$, and $T_2^*$ under two acceleration rates (R = 6.5 and R = 11.8). (A) Boxplot about the scan-rescan repeatability CV of $T_1$ using 3D-QALAS and MIMOSA. (B) Boxplot about the scan-rescan repeatability CV of $T_2$ using 3D-QALAS and MIMOSA. (C) Boxplot about the scan-rescan repeatability CV of $T_2^*$ using MIMOSA.

Figure 8 presents box plots of repeatability CV for $T_1$, $T_2$, and $T_2^*$ maps at R = 6.5 and 11.8, evaluating repeatability by analyzing nine ROIs from each of the two subjects. Notably, all maps acquired at R = 11.8 exhibit both a narrower interquartile range and a lower median of CV, indicating improved repeatability for both 3D-QALAS and MIMOSA. Furthermore, MIMOSA demonstrates slightly improved repeatability in both $T_1$ and $T_2$ mapping compared to 3D-QALAS, with higher ICC values for the repeatability CV at R = 11.8 ($T_1$, MIMOSA and 3D-QALAS: 0.998 and 0.993; $T_2$, MIMOSA and 3D-QALAS: 0.973 and 0.969; $T_2^*$, MIMOSA: 0.947; QSM, MIMOSA: 0.992; $\chi_{para}$, MIMOSA: 0.987; $\chi_{dia}$, MIMOSA: 0.977). Figure S11 shows all scan-rescan quantitative maps from five subjects.

## 3.3.3 Mesoscale quantitative mapping

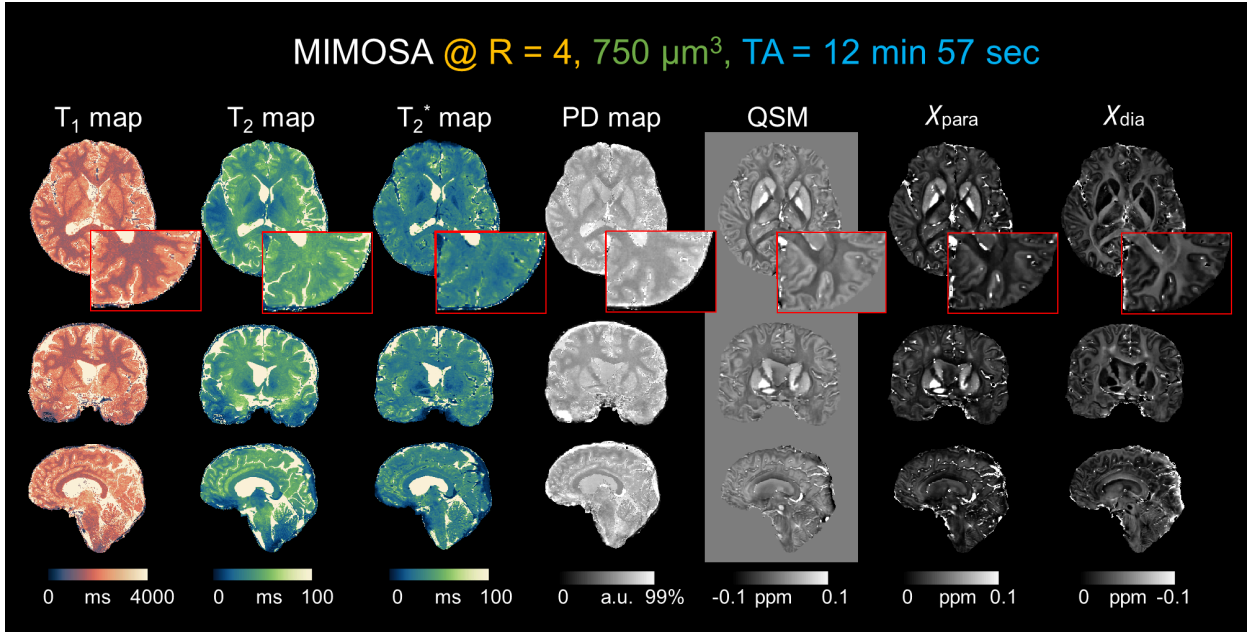

FIGURE 9. Simultaneous whole-brain $T_1$, $T_2$, $T_2^*$, PD, QSM, para- and diamagnetic susceptibility mapping at 750 μm³ resolution in 13 minutes at 7T using MIMOSA.

Figure 9 depicts the whole-brain mesoscale $T_1$, $T_2$, $T_2^*$, PD, QSM, and para- and diamagnetic susceptibility maps at the isotropic resolution of 750 μm, demonstrating the mesoscale acquisition capability of MIMOSA.

# 4. DISCUSSION AND CONCLUSION

We proposed MIMOSA to enable simultaneous $T_1$, $T_2$, $T_2^*$, PD, QSM, para- and diamagnetic susceptibility mapping. Complementary variable-density spiral-like sampling trajectory was combined with FLASH and MGRE modules to enable highly efficient acquisition. Simulations were performed to further optimize the sequence. MZS-SSL was used to reconstruct the acquired multi-contrast k-space data, demonstrating robust estimation accuracy for $T_1$, $T_2$, and $T_2^*$ mapping across three different acceleration rates (R = 3.3, 6.5, and 11.8). The accuracy of parameter estimation using MIMOSA was validated in both phantom and in-vivo experiments by comparison with 3D-QALAS and reference methods. The scan-rescan experiments show that MIMOSA has high repeatability with the ICC of 0.998, 0.973, 0.947, 0.992, 0.987, and 0.977 for $T_1$, $T_2$, $T_2^*$, QSM, $\chi_{para}$, and $\chi_{dia}$, respectively. Simultaneous whole-brain $T_1$, $T_2$, $T_2^*$, PD, and source separation QSM was achieved with 1 mm isotropic resolution in 3 minutes at 3T, and with 750 μm isotropic resolution in 13 minutes at 7T.

The parameter estimation accuracy of MIMOSA was evaluated through simulation studies and compared with 3D-QALAS. Building on 3D-QALAS, MIMOSA provides additional quantitative parameters, including $T_2^*$, QSM, para- and diamagnetic susceptibility maps, at the expense of a 34% longer TR (6.03 s vs. 4.5 s) due to the additional MGRE module. However, it remains faster than acquiring these maps separately using 3D-QALAS and an additional 3D-MGRE scan (which would have required 3 min 52 s at R = 11.8 vs. the current 2 min 55 s at R = 11.8 with MIMOSA). The MGRE module features fully programmable acquisition parameters, including adjustable echo time and configurable echo train length, enabling protocol optimization for applications across field strengths and tissue types. Simulations were conducted to optimize the sequence by comparing six different MIMOSA configurations (Table S1) against 3D-QALAS. The simulation results indicated that MIMOSA provides more accurate estimation of $T_1$ and $T_2$ maps with reduced NRMSE (Figure 2). Further investigation of different MIMOSA configurations (Table S1) reveals a trade-off between acquisition efficiency and parameter estimation accuracy in tissues with long $T_1$ relaxation times. Among them, the most efficient configuration (Configuration 3.2 in Table S1) demonstrated parameter estimation accuracy comparable to both the most accurate configuration (Configuration 1.1 in Table S1) and 3D-QALAS, and was therefore selected as the final setup for phantom and in-vivo experiments. Relatively larger estimation error was observed in CSF (Figure 2, Figure S5, and Table S1) due to its longer relaxation times compared to parenchymal tissues. Since CSF is rarely a diagnostic focus in qMRI, we prioritized accuracy for brain tissue parameters, where pathological changes are most relevant. Nevertheless, MIMOSA's flexibility in adjusting the TE of $T_2$ preparation pulse, inversion time, and TE of MGRE module provides potential for capturing atypical relaxation time evolution. If CSF quantification is desired, the inversion time can be optimized accordingly, though at the potential cost of increased scan time.

The feasibility of MIMOSA was validated by comparing MIMOSA with 3D-QALAS in both phantom and in-vivo experiments. In the phantom experiments, MIMOSA showed slightly improved linearity and accuracy for $T_1$ and $T_2$ estimation, with regression slopes of 0.9262 and 1.1714, and $R^2$ of 0.9971 and 0.9976, respectively, compared to 3D-QALAS (slopes: 0.8908 and 1.3689, $R^2$: 0.9964 and 0.9904). These improvements may be attributed to the $T_2^*$ mapping as a prior for joint estimation of $T_1$, $T_2$, $T_2^*$, and PD. MZS-SSL was employed to facilitate fast acquisition by leveraging information across contrasts and slices for the reconstruction of highly undersampled MIMOSA data. Benefiting from acquisition of multi-echo images from the MGRE module with complimentary sampling, MIMOSA was demonstrated to have good reconstruction quality with reduced noise in both $T_1$ and $T_2$ mapping of the in-vivo experiments. The results showed that up to R = 11.8-fold acceleration factor can be achieved without noticeably compromising parameter estimation accuracy.

The repeatability of 3D-QALAS has been previously investigated,[22] and in scan-rescan experiments, MIMOSA demonstrated a slightly higher ICC for repeatability CV ($T_1$: 0.998 vs. 0.993; $T_2$: 0.973 vs. 0.969 for MIMOSA and 3D-QALAS, respectively). Notably, the scan-rescan experiments demonstrated enhanced repeatability for MIMOSA at R = 11.8 compared to R = 6.5, a consistent trend that was also observed in 3D-QALAS. This improvement may be attributed to the reduced physiological fluctuations and involuntary motion resulting from the shorter scan time. Building on this observed improvement in repeatability, future work could focus on further enhancing k-space sampling efficiency and reducing scan time. Since MIMOSA is currently encoded by Cartesian readouts, incorporating non-Cartesian sampling trajectories, such as spiral and radial sampling, into the readouts could be a promising direction. Further studies with larger cohorts can help assess the generalizability of findings and facilitate robust clinical validation.

Mesoscale quantitative $T_1$, $T_2$, $T_2^*$, PD, QSM, para- and diamagnetic susceptibility mapping with 750 μm isotropic resolution in 13 min was achieved at 7T using MIMOSA. By using the same acceleration rate of R = 11.8 in the 3T experiments, a scan time of approximately 4.4 min at 7T is possible. Compared to the results at 1 mm resolution at 3T, mesoscale quantitative mapping offers enhanced resolution and improved SNR, providing detailed insights into tissue composition and magnetic properties. This is particularly beneficial for studying fine cortical structures[66] and detecting smaller lesions at early stages.[67] To mitigate the effects of increased $B_1^+$ inhomogeneity at 7T, we prolonged the length of the adiabatic refocusing pulses used in the $T_2$ sensitizing phase from 3 ms to 8 ms to boost $B_1^+$ robustness.[36] However, the shading artifacts resulting from inhomogeneity-induced signal loss remain observable in the $T_2$ maps at 7T (Figure 8). These artifacts can be further mitigated through the implementation of parallel transmission (pTx)[68] and deep learning-based parameter estimation methods.[36,69,70] While mesoscale quantitative mapping provides superior spatial resolution, the typically prolonged scan time makes it vulnerable to motion artifacts. Current implementation of MIMOSA employs a variable-density spiral-like Cartesian k-space sampling trajectory, which acquires low-frequency k-space data in each TR. This sampling strategy inherently enables self-navigated motion correction through transformation matrices derived from central k-space signals, which can be integrated into the reconstruction pipeline as demonstrated by Fujita et al.[71], but this capability was not used for the results presented herein.

MIMOSA achieved magnetic susceptibility source separation from a single, multi-parametric scan. Existing multi-parametric mapping techniques[25–27] have been proposed to provide $T_1$, $T_2$, $T_2^*$, and QSM simultaneously. However, these methods face several limitations including long scan times, limited through plane resolution, or residual artifacts due to undersampling. For instance, 3D qRF-MRF[27] provides simultaneous quantification of $T_1$, $T_2$, $T_2^*$, and QSM at a resolution of $1.2 \times 1.2 \times 5$ mm$^3$ in 5 min 30 s. Additionally, while these techniques could potentially disentangle the para- and diamagnetic concentrations from the bulk susceptibility, this application remains unexplored. As $R_2'$-based susceptibility source separation necessitates $R_2'$ mapping to complement frequency shift information, which requires simultaneous acquisition of both $T_2$ and $T_2^*$ maps that are typically obtained from MGRE and spin-echo (SE) sequences, respectively. This approach suffers from long scan times and potential misalignment between the sequences. MIMOSA overcomes these challenges by enabling simultaneous acquisition of both $T_2$ and $T_2^*$ maps, allowing for more efficient source separation without the limitations associated with traditional acquisition methods. $\chi$-sepnet,[31] a deep learning model trained on multi-orientation head data to effectively suppress streaking artifacts, was employed for QSM and susceptibility source separation. Since the relaxometric constant required for susceptibility source separation at 7T has not been experimentally established,[72] we employed the empirically determined value of 114 Hz/ppm from 3T studies.[31] This approach, while practical, may introduce biases in susceptibility source separation at 7T due to field strength-dependent variations. However, these potential limitations can be addressed through linear regression analysis of multiple gray matter ROIs exhibiting predominantly paramagnetic susceptibility characteristics,[10] enabling the estimation of 7T relaxometric constant for improved accuracy.

MIMOSA was developed using Pulseq and can be readily deployed across various scanners. Pulseq is a vendor-neutral pulse sequence development platform and has been applied to develop numerous sequences, such as the standard CEST[73] and EPI based diffusion MRI[74]. In Pulseq, sequences can be executed efficiently using vendor-specific interpreters in a standardized manner, making it highly suitable for multi-center studies.[75,76] This study illustrated its use across different scanner software baselines on Siemens platforms (VE11 and XA30 at 3T, and VE12 at 7T). The application of MIMOSA in multi-center studies can be investigated in the future.

One limitation of MIMOSA is the reconstruction time. The training of the MZS-SSL model required about 18 hours using a single GPU. However, by utilizing this pre-trained network and implementing transfer learning for new subject data, the reconstruction time can be significantly reduced to approximately 3 hours. Further acceleration of model training can be explored through multiple approaches: (1) optimization of the conjugate gradient (CG) algorithm-based data consistency layer through implementation of preconditioning matrices (e.g., diagonal or low-rank approximations) to enhance CG convergence rates; (2) development of knowledge distillation frameworks, where a computationally efficient student model is trained under the guidance of a larger teacher model to maintain reconstruction accuracy while significantly reducing computational demands; (3) implementation of distributed training paradigms utilizing multi-GPU architectures to parallelize computational workloads. Additionally, future exploration of online reconstruction implementation could substantially improve workflow efficiency of MIMOSA. This could involve deployment of the pre-trained model on a remote server integrated with the Framework for Image Reconstruction (FIRE),[77] enabling data transmission and retrieval between the scanner and the remote computational resources.[78,79]

In MIMOSA, parameters are estimated by dictionary matching based on a single-component relaxation model. The dictionary was generated under the assumption that image contrast is primarily determined by the echo sampled at the k-space center. However, this assumption neglects relaxation effects during the echo train, which can introduce spatial blurring and biases in quantitative mapping. Low-rank subspace reconstruction[24,56,80–82] can be employed to address this problem by using subspace bases to capture the complete signal evolution throughout the echo train. Moreover, while a single-component relaxation model is generally adequate for most neuroimaging applications where tissue water dominates the signal, dedicated fat/water separation may be needed for accurate quantification in fat-rich anatomical regions.

Another challenge of $T_2^*$ mapping with MIMOSA is the sensitivity to $B_0$ field inhomogeneity and signal evolution dynamics. As the MGRE module is applied after the adiabatic IR pulse, $T_2^*$ mapping is vulnerable to the initial signal state, leading to overestimation of $T_2^*$ mapping in the regions with poor $B_0$ uniformity (orange arrows in Figure S12). To mitigate this problem, an out-to-center sampling order was applied for the MGRE module, allowing for enough recovery time before sampling at the k-space center. Figure S12 compares the $T_2^*$ mapping using a center-out and an out-to-center sampling order, respectively. By using an out-to-center sampling order for the MGRE module, the overestimation of $T_2^*$ was corrected. However, there exists some systematic underestimation of $T_2^*$ caused by macroscopic field inhomogeneity at air/tissue interfaces. This phenomenon was consistently observed in $T_2^*$ mapping of both 3D-MGRE sequence and MIMOSA, as indicated by red arrows in Figure S12. Advanced post processing correction techniques[83,84] can be implemented to address this problem.

In conclusion, we proposed a novel multi-parametric mapping technique, MIMOSA. Highly efficient acquisition was achieved by carefully designing the sampling strategy and sequence optimization based on simulations. Highly undersampled k-space data with up to R = 11.8-fold acceleration was jointly reconstructed using MZS-SSL, achieving comparable or superior parameter estimation accuracy and improved scan-rescan repeatability compared to 3D-QALAS. Whole-brain $T_1$, $T_2$, $T_2^*$, PD, QSM, para- and diamagnetic susceptibility mapping were obtained with 1 mm isotropic resolution in 3 min at 3T and 750 µm isotropic resolution in 13 min at 7T. MIMOSA demonstrated potential for highly-efficient quantitative imaging, offering detailed insights into brain composition.

**Data Availability Statement**

The source code of MIMOSA, including sequence Pulseq files and reconstruction algorithms, is available at https://github.com/yutingchen11/MIMOSA.

# Supplement Information

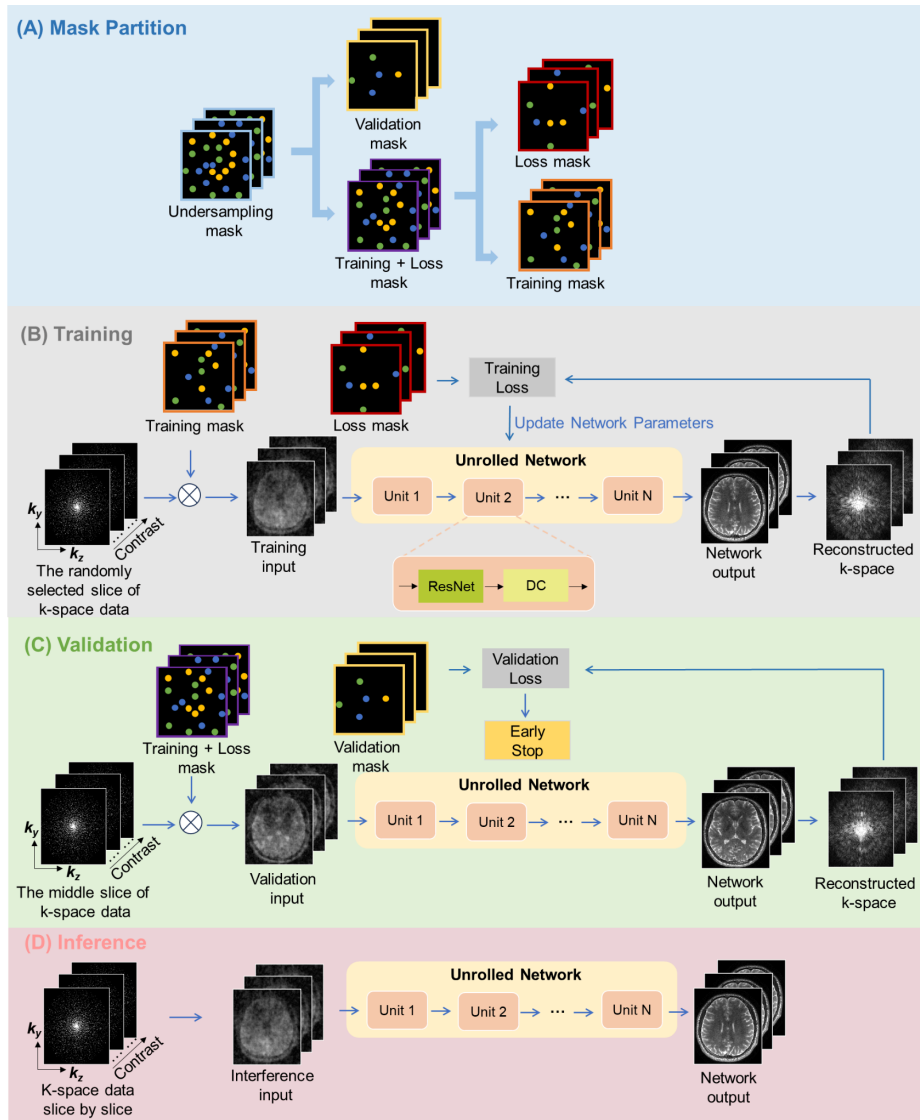

Supporting Information Figure S1. An overview of multi-contrast/-slice ZS-SSL. (A) The undersampling k-space masks are partitioned into three disjoint subsets: validation, loss, and training masks. (B) The slices from all contrasts are randomly shuffled, and one slice is selected and multiplied with the generated training and loss k-space mask. Subsequently, the k-space data is converted into images where different contrasts are incorporated into the channel dimension, serving as input to the unrolled network. The loss mask is used to compute the training loss and update the network parameters. (C) Validation is conducted in each training epoch as a stopping criterion by computing the validation loss of the middle slice using the validation mask. (D) After training, whole-brain volume with multi-contrast can be reconstructed slice by slice. DC: data consistency layer.

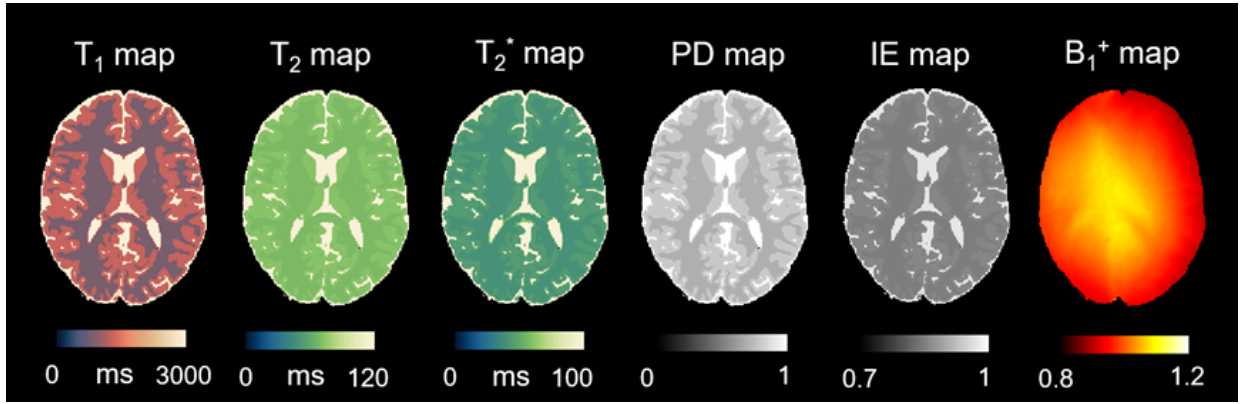

Supporting Information Figure S2. The generated reference $T_1$, $T_2$, $T_2^*$, PD, IE, and $B_1^+$ map used for simulations. PD: proton density. IE: inversion efficiency. $B_1^+$: radio frequency transmit field map.

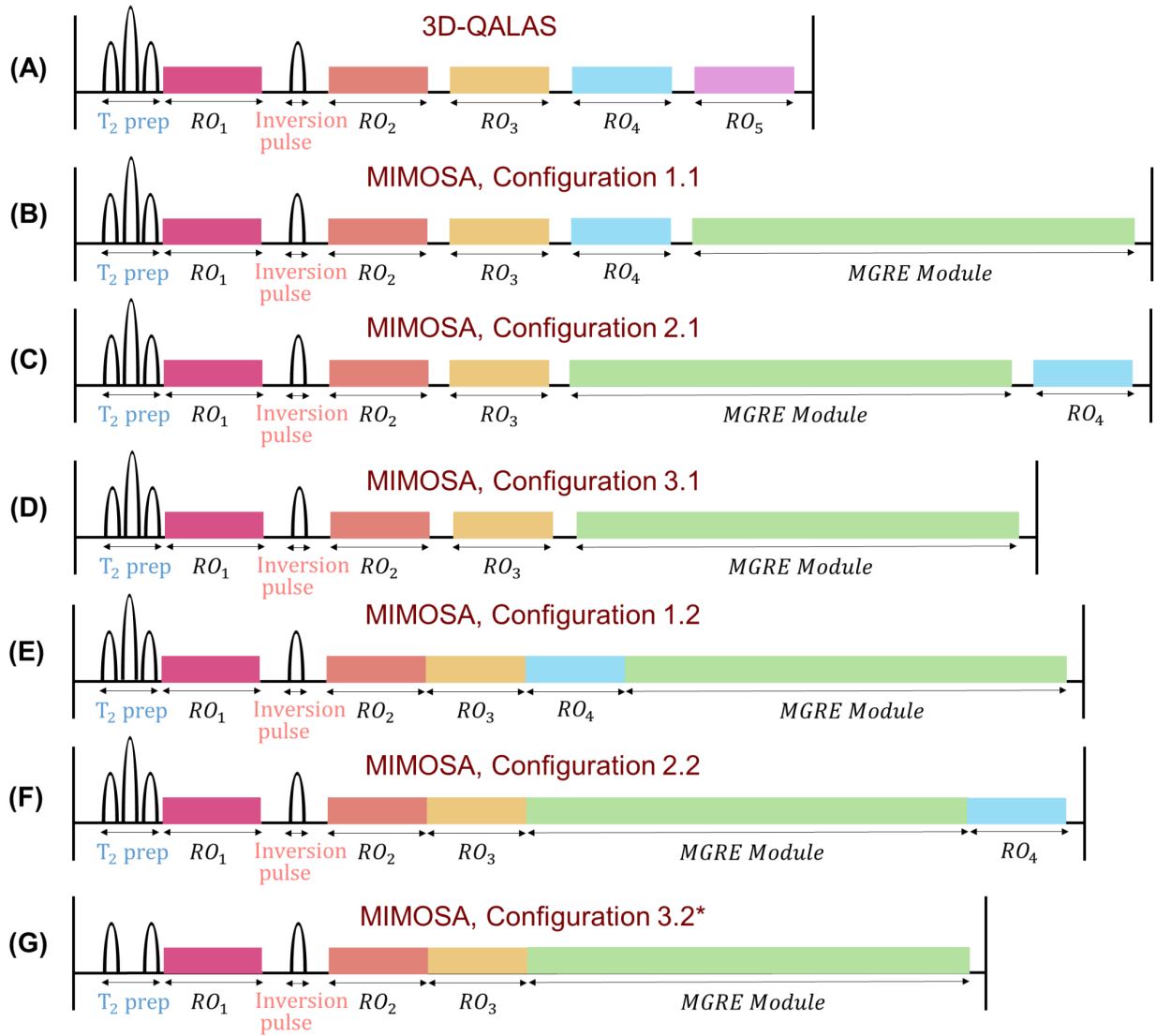

Supporting Information Figure S3. The sequence diagram of (A) 3D-QALAS and MIMOSA with three different configurations with (B-D) and without delays (E-G) between FLASH readouts. The final used configuration of MIMOSA was marked with *. RO: 3D Fast Low Angle Shot (FLASH) readout. MGRE Module: 3D multi-echo gradient echo module. 3D-QALAS: 3D-quantification using an interleaved Look-Locker acquisition sequence with $T_2$ preparation pulse. MIMOSA: Multi-parametric and Multi-echo Imaging and Mapping with Optimized Simultaneous Acquisition.

Supporting Information Table S1. The normalized root mean squared error (NRMSE) of $T_1$, $T_2$, $T_2^*$, and PD mapping across three tissue types (WM, GM, and CSF) for 3D-QALAS and MIMOSA with the configurations of Figure S3.

| Sequence | Tissue | NRMSE (%) | | | |
|---|---|---|---|---|---|
| | | $T_1$ | $T_2$ | PD | $T_2^*$ |
| **3D-QALAS** 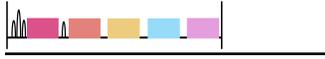 | WM | 1.69 | 1.46 | 0.82 | N/A |
| | GM | 2.02 | 1.44 | 1.22 | N/A |
| | CSF | 5.22 | 18.55 | 4.96 | N/A |
| **MIMOSA Config 1.1** 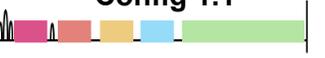 | WM | 1.16 | **1.23** | 0.66 | 1.41 |
| | GM | **1.02** | **1.10** | 0.65 | 1.41 |
| | CSF | **2.22** | 8.03 | **1.94** | **5.54** |
| **MIMOSA Config 2.1** 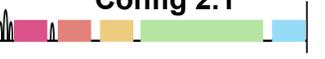 | WM | 1.12 | 1.29 | **0.59** | 1.41 |
| | GM | 1.18 | 1.20 | **0.59** | 1.39 |
| | CSF | 3.20 | 9.33 | 2.44 | 5.78 |
| **MIMOSA Config 3.1** 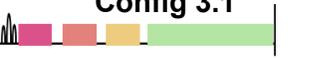 | WM | 1.18 | **1.23** | 0.66 | **1.40** |
| | GM | 1.18 | 1.12 | 0.65 | **1.38** |
| | CSF | 3.08 | 8.42 | 2.43 | 5.92 |
| **MIMOSA Config 1.2** 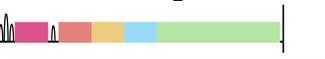 | WM | **1.09** | 1.24 | 0.66 | **1.40** |
| | GM | 1.03 | 1.11 | 0.66 | 1.41 |
| | CSF | 3.01 | **7.27** | 2.39 | 5.72 |
| **MIMOSA Config 2.2** 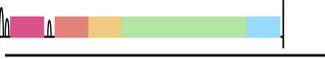 | WM | 1.14 | 1.37 | **0.59** | 1.41 |
| | GM | 1.27 | 1.24 | 0.61 | 1.42 |
| | CSF | 3.72 | 9.84 | 2.79 | 6.04 |
| **MIMOSA Config 3.2*** 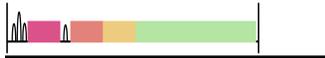 | WM | 1.17 | **1.23** | 0.66 | 1.41 |
| | GM | 1.29 | 1.13 | 0.66 | 1.41 |
| | CSF | 3.47 | 8.58 | 2.65 | 5.94 |

The final proposed configuration was marked with * . 3D-QALAS: 3D-quantification using an interleaved Look-Locker acquisition sequence with $T_2$ preparation pulse. MIMOSA: Multi-parametric and Multi-echo Imaging and Mapping with Optimized Simultaneous Acquisition. Config: configuration; PD: proton density. WM: white matter; GM: gray matter; CSF: cerebrospinal fluid.

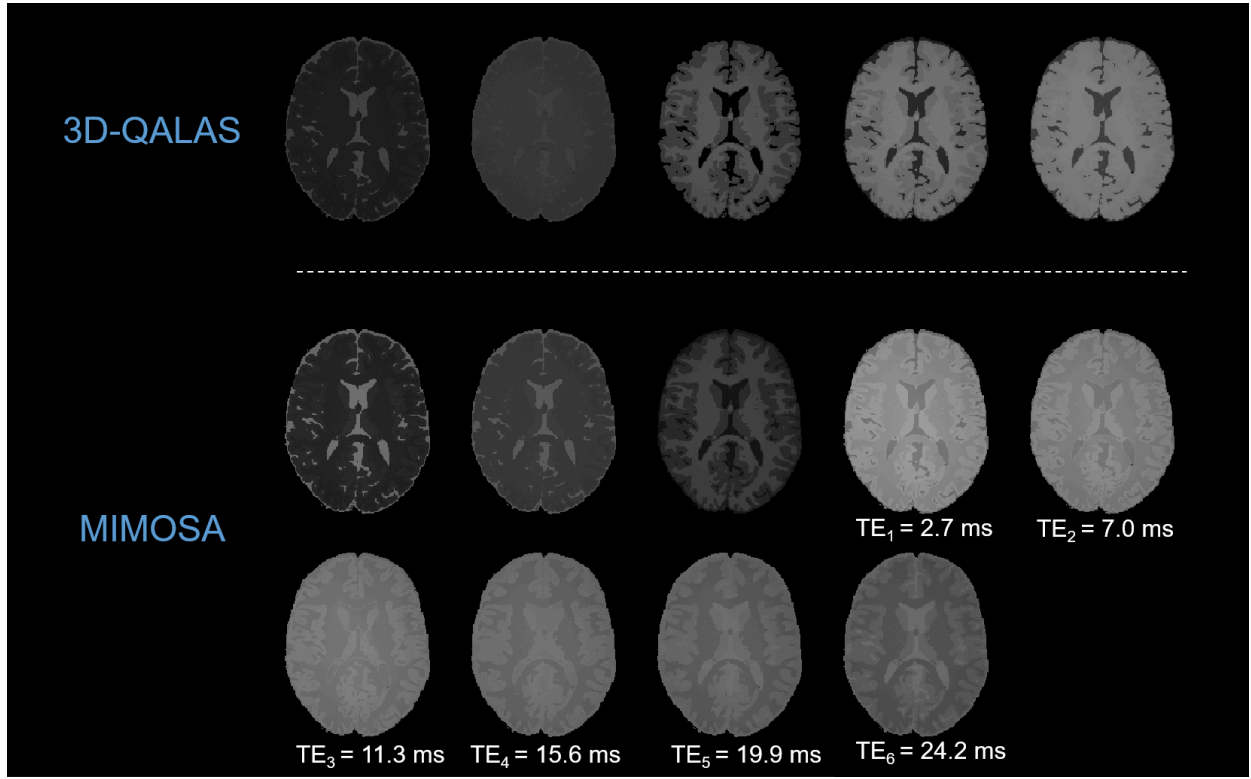

Supporting Information Figure S4. The simulated multi-contrast images of 3D-QALAS and MIMOSA. 3D-QALAS: 3D-quantification using an interleaved Look-Locker acquisition sequence with $T_2$ preparation pulse. MIMOSA: Multi-parametric and Multi-echo Imaging and Mapping with Optimized Simultaneous Acquisition.

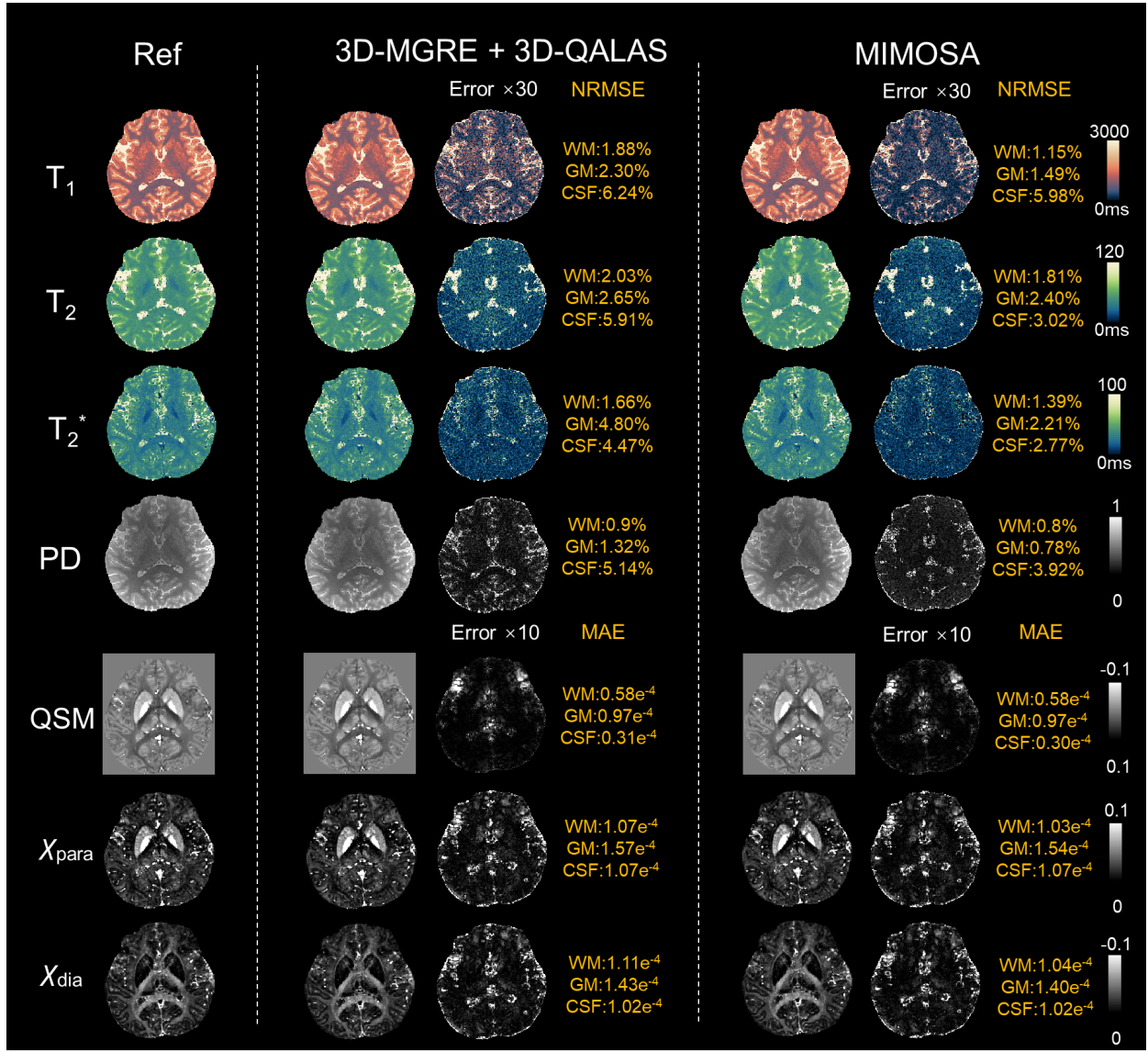

Supporting Information Figure S5. Results of numerical phantom simulations. The reference $T_1$, $T_2$, and PD maps were acquired using 3D-QALAS. The reference $T_2^*$, QSM, and susceptibility source separation maps were obtained using 3D-MGRE, where the $T_2$ map from 3D-QALAS was used for source separation. These reference maps were used to simulate 3D-MGRE, 3D-QALAS, and MIMOSA. The quantitative parameter maps (second and fourth column) and absolute error maps (magnified by a factor of 30 or 10) were calculated by comparing the estimated parameter maps with the reference maps. The corresponding NRMSEs and MAE were also provided. 3D-MGRE: 3D multi-echo gradient echo; 3D-QALAS: 3D-quantification using an interleaved Look-Locker acquisition sequence with $T_2$ preparation pulse; MIMOSA: Multi-parametric and Multi-echo Imaging and Mapping with Optimized Simultaneous Acquisition. NRMSE: normalized root mean squared error; MAE: mean absolute error.

Supporting Information Table S2. Protocols of IR-FSE, SE-FSE, and 3D-MGRE sequences used in phantom experiments.

| | IR-FSE | SE-FSE | 3D-MGRE |
|---|---|---|---|
| FOV | $200 \times 200 \ mm^2$ | $200 \times 200 \ mm^2$ | $240 \times 224 \times 192 \ mm^3$ |
| Matrix Size | $200 \times 200$ | $200 \times 200$ | $240 \times 224 \times 48$ |
| Slice Thickness | 4 mm | 4 mm | 4 mm |
| TI | [35, 100, 150, 250, 500, 1000, 2000, 3000, 4000] ms | - | - |
| TE | 9 ms | [10, 30, 50, 70, 90, 120, 200, 300, 400] ms | [2.7, 7.0, 11.3, 15.6, 19.9, 24.2] ms |
| TR | 8200 ms | 1500 ms | 27.5 ms |
| Acceleration Rate | 2 | 2 | 1 |
| Scan Time | 21 min 14 s | 20 min 42 s | 5 min 2 s |

IR-FSE: inversion-recovery fast-spin-echo; SE-FSE: single-echo fast-spin-echo; 3D-MGRE: 3D multi-echo gradient echo.

Supporting Information Table S3. Protocols of IR-FSE, SE-FSE, and 3D-MGRE sequences used in in-vivo experiments.

| | IR-FSE | SE-FSE | 3D MGRE |
|---|---|---|---|
| FOV | $192 \times 192\ mm^2$ | $192 \times 192\ mm^2$ | $240 \times 224 \times 192\ mm^3$ |
| Matrix Size | $192 \times 192$ | $192 \times 192$ | $240 \times 224 \times 192$ |
| Slice Thickness | 3 mm | 3 mm | 1 mm |
| TI | [100, 300, 600, 800, 1000, 2000, 3500] ms | - | - |
| TE | 9 ms | [25, 50, 75, 100, 125, 150, 200] ms | [2.7, 7.0, 11.3, 15.6, 19.9, 24.2] ms |
| TR | 6000 ms | 800 ms | 27.5 ms |
| Acceleration Rate | 2 | 2 | 3.3 |
| Scan Time | 8 min 38 s | 10 min 23 s | 6 min 10 s |

IR-FSE: inversion-recovery fast-spin-echo; SE-FSE: single-echo fast-spin-echo; 3D-MGRE: 3D multi-echo gradient echo.

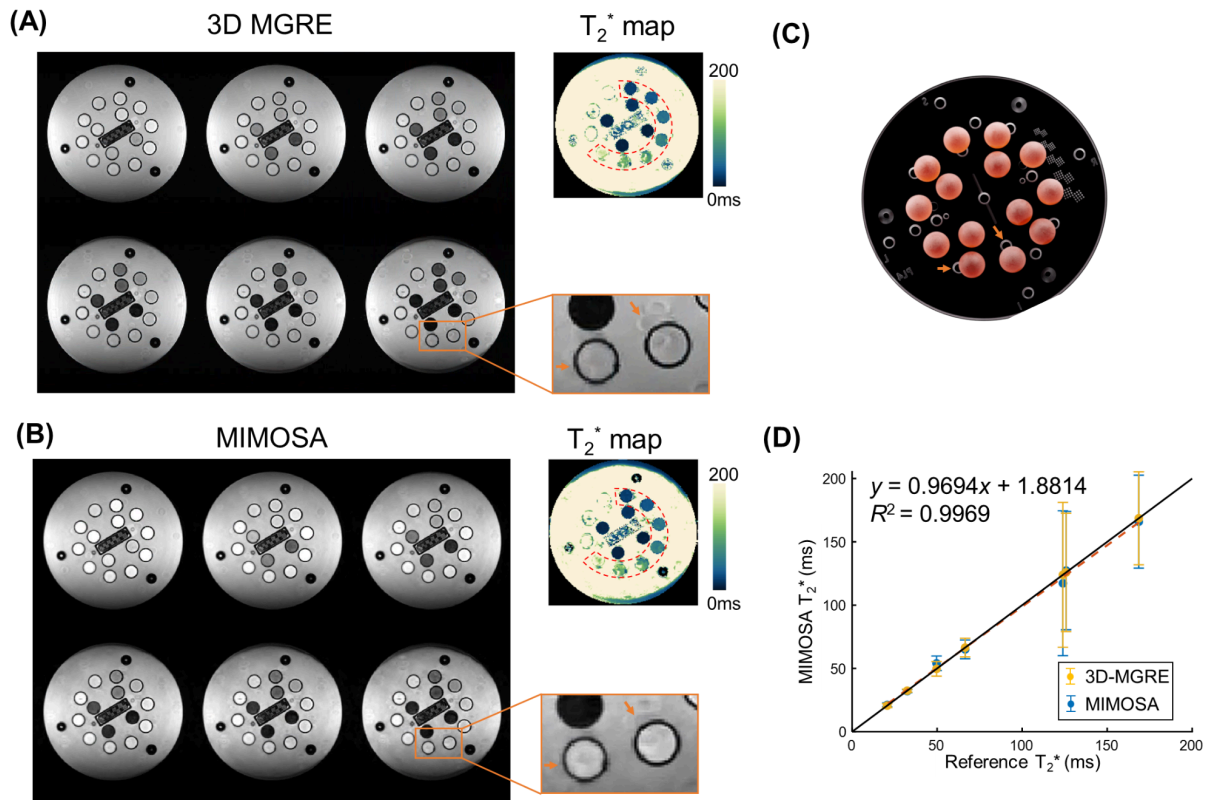

**Supplement Information Figure S6**. Comparison of $T_2^*$ mapping using 3D-MGRE and MIMOSA. The multi-contrast images of 3D MGRE (A) and multi-contrast module of MIMOSA (B) were used for estimation of $T_2^*$. (C) The image of phantom structure[1] (orange arrows) that induces dipole-like artifacts (orange zoom box) of both (A) and (B). (D) Regression analysis of $T_2^*$ values measured within the regions of interest (ROIs, red dotted lines). [1] https://qmri.com/product/premium-system-phantom/

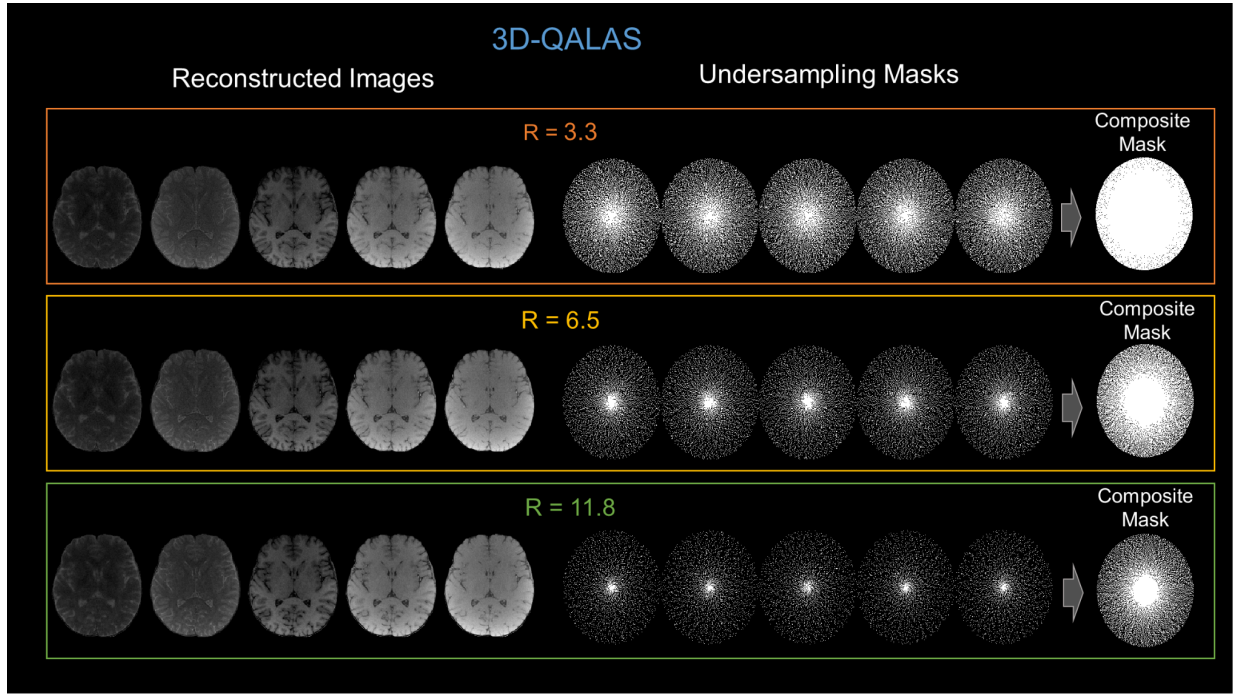

Supporting Information Figure S7. Reconstructed multi-contrast 3D-QALAS images and the corresponding undersampling masks at R = 3.3, 6.5, and 11.8, respectively. The composite mask, obtained by combining the complementary undersampling masks across all contrasts, is presented in the last column. R: acceleration rate. 3D-QALAS: 3D-quantification using an interleaved Look-Locker acquisition sequence with $T_2$ preparation pulse. MIMOSA: Multi-parametric and Multi-echo Imaging and Mapping with Optimized Simultaneous Acquisition. R: acceleration rate.

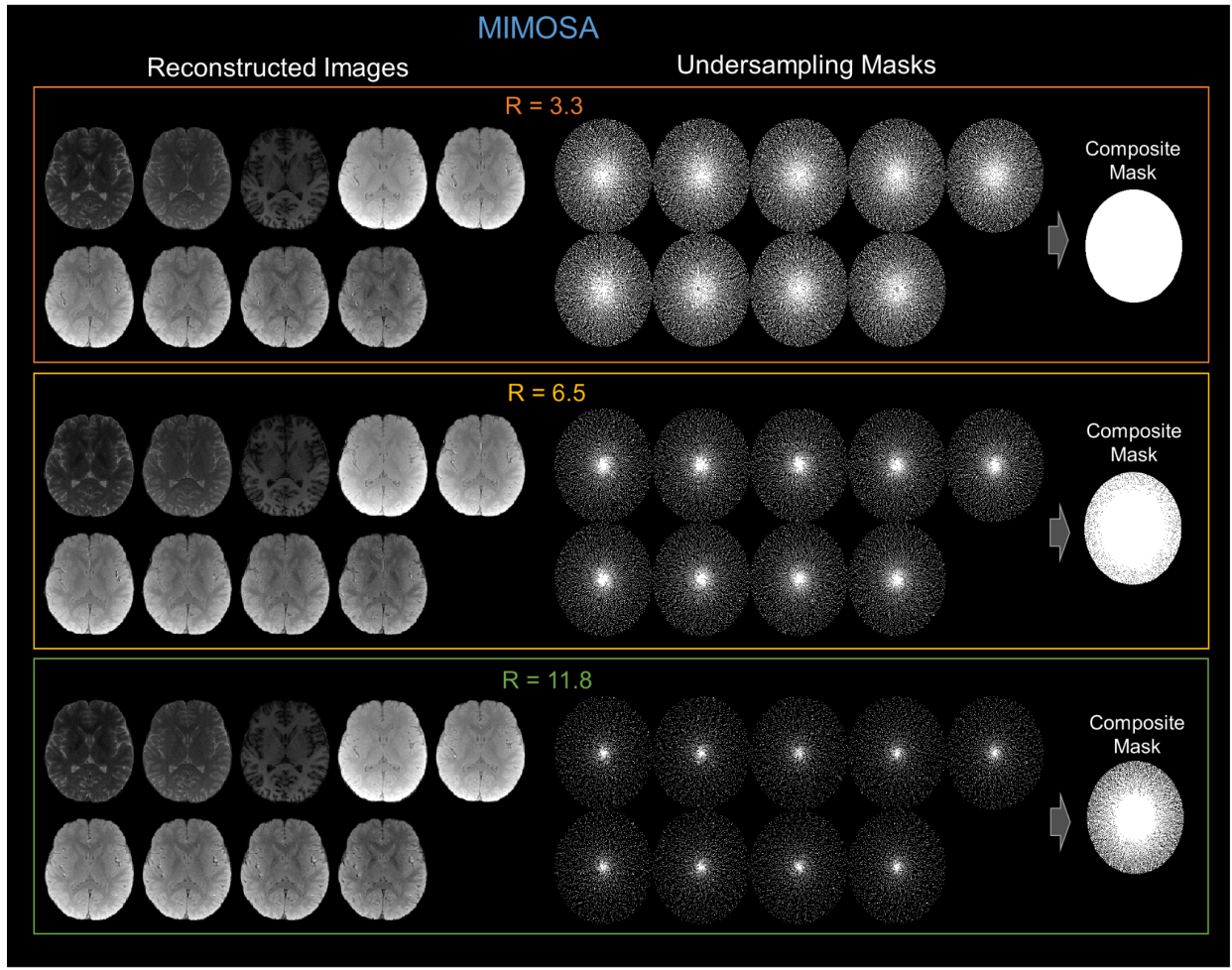

Supporting Information Figure S8. Reconstructed multi-contrast MIMOSA images and the corresponding undersampling masks at R = 3.3, 6.5, and 11.8, respectively. The composite mask, obtained by combining the complementary undersampling masks across all contrasts, is presented in the last column. R: acceleration rate. 3D-QALAS: 3D-quantification using an interleaved Look-Locker acquisition sequence with $T_2$ preparation pulse. MIMOSA: Multi-parametric and Multi-echo Imaging and Mapping with Optimized Simultaneous Acquisition. R: acceleration rate.

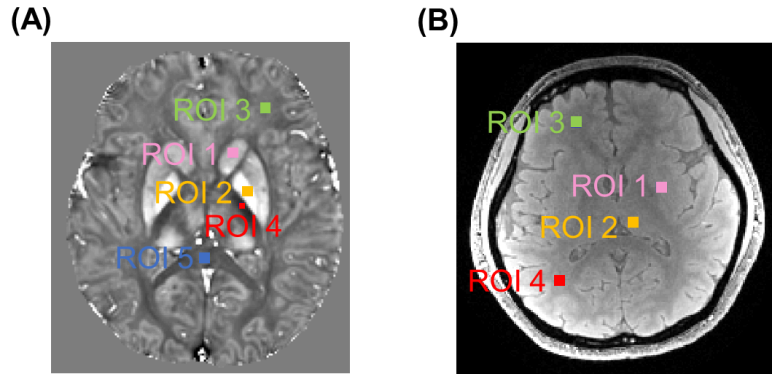

Supporting Information Figure S9. Region of interests (ROIs) used for quantitative analysis of QSM, para- and diamagnetic susceptibility mapping (A), and $T_1$ and $T_2$ mapping of 3D-QALAS and MIMOSA with matched scan time (B).

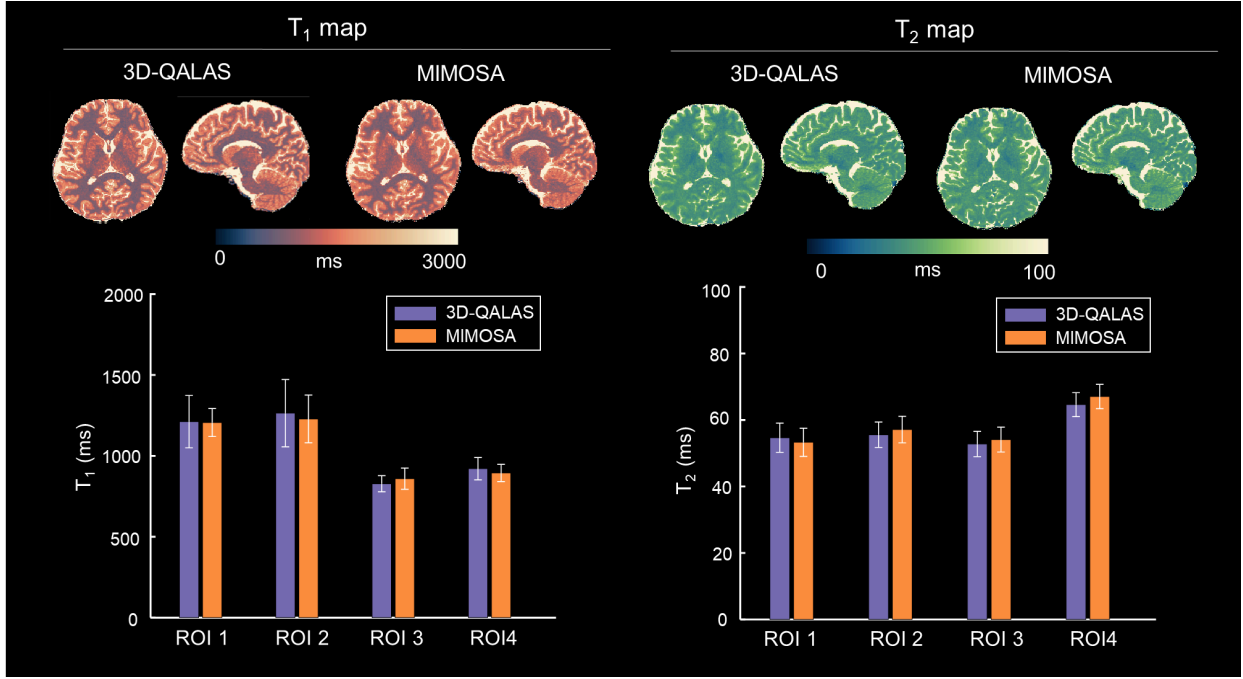

Supporting Information Figure S10. Comparisons of $T_1$ and $T_2$ mapping using 3D-QALAS (R = 8.7) and MIMOSA (R = 11.8) at matched scan times (2 min 55 sec). Four ROIs were selected for quantitative comparison.

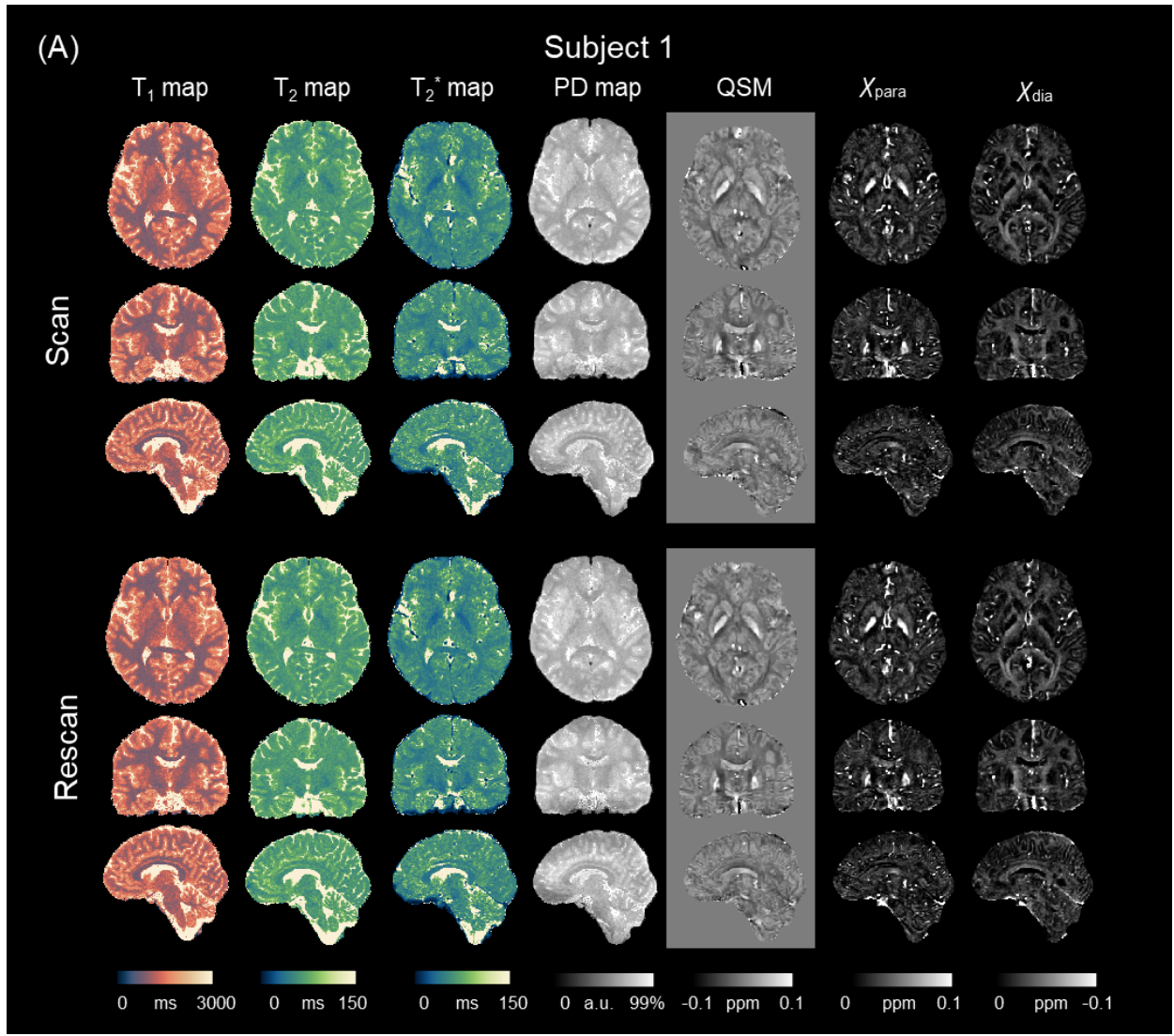

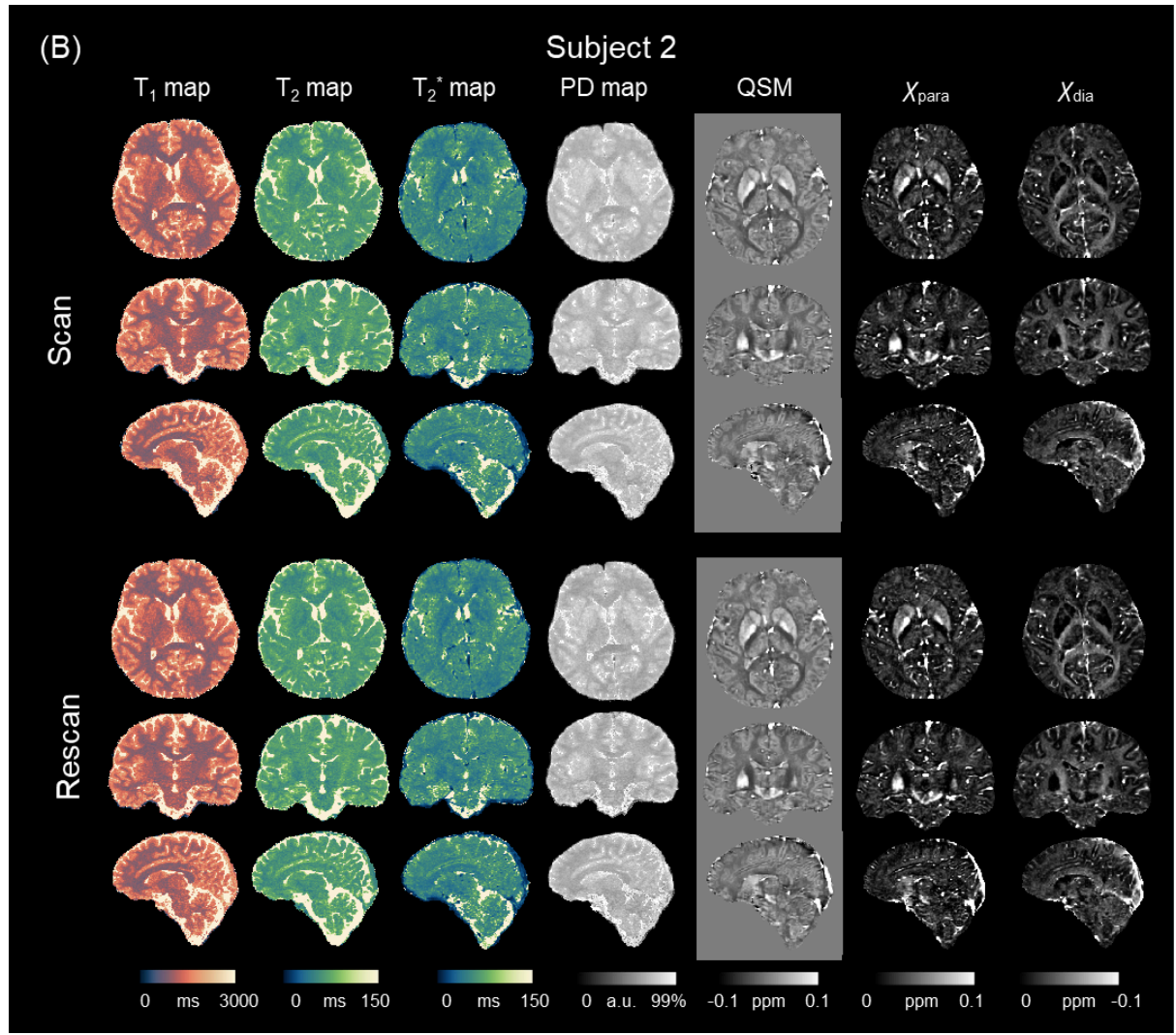

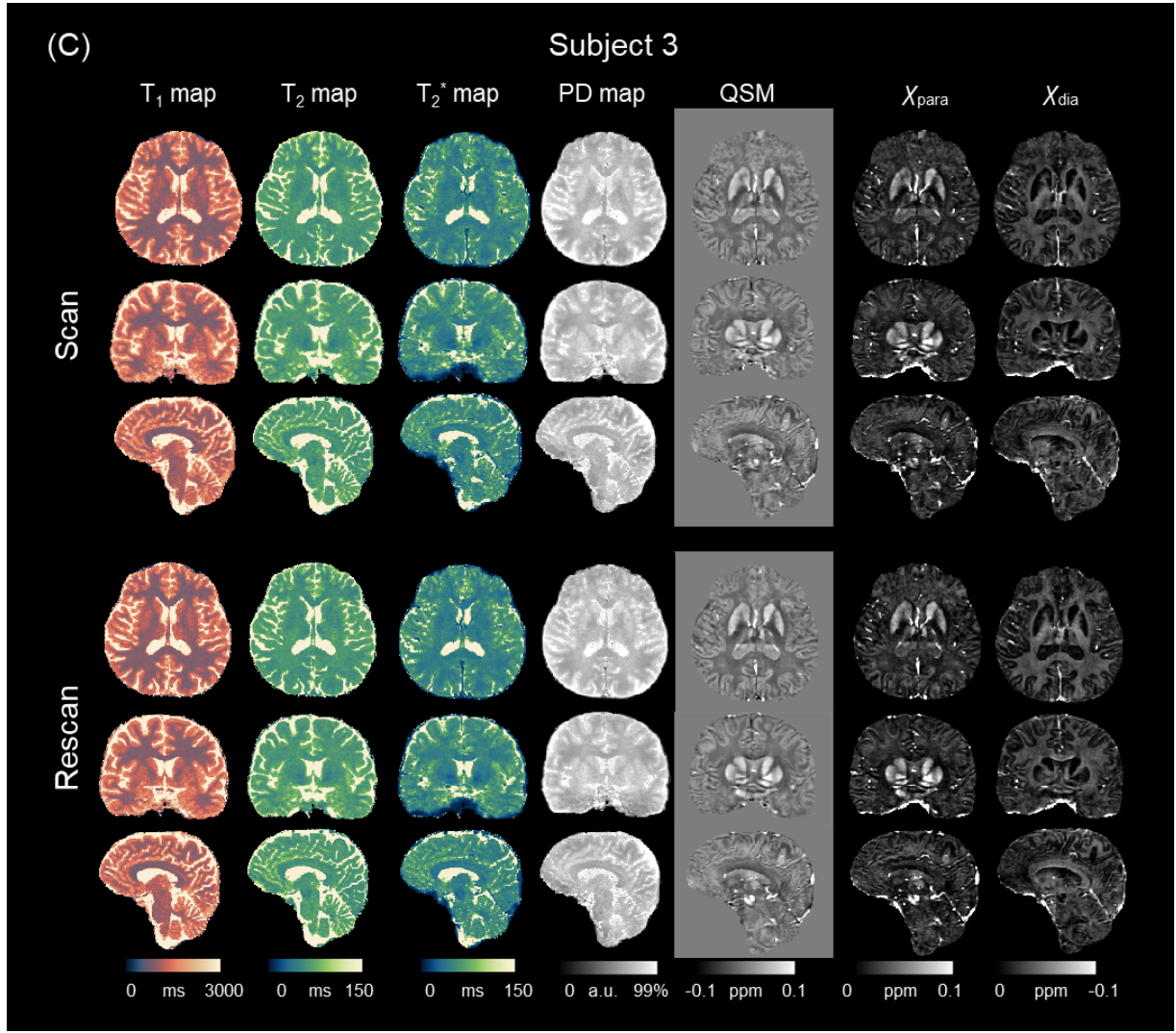

(C)

Subject 3

T₁ map   T₂ map   T₂* map   PD map   QSM   $X_{para}$   $X_{dia}$

Scan

Rescan

0 ms 3000   0 ms 150   0 ms 150   0 a.u. 99%   -0.1 ppm 0.1   0 ppm 0.1   0 ppm -0.1

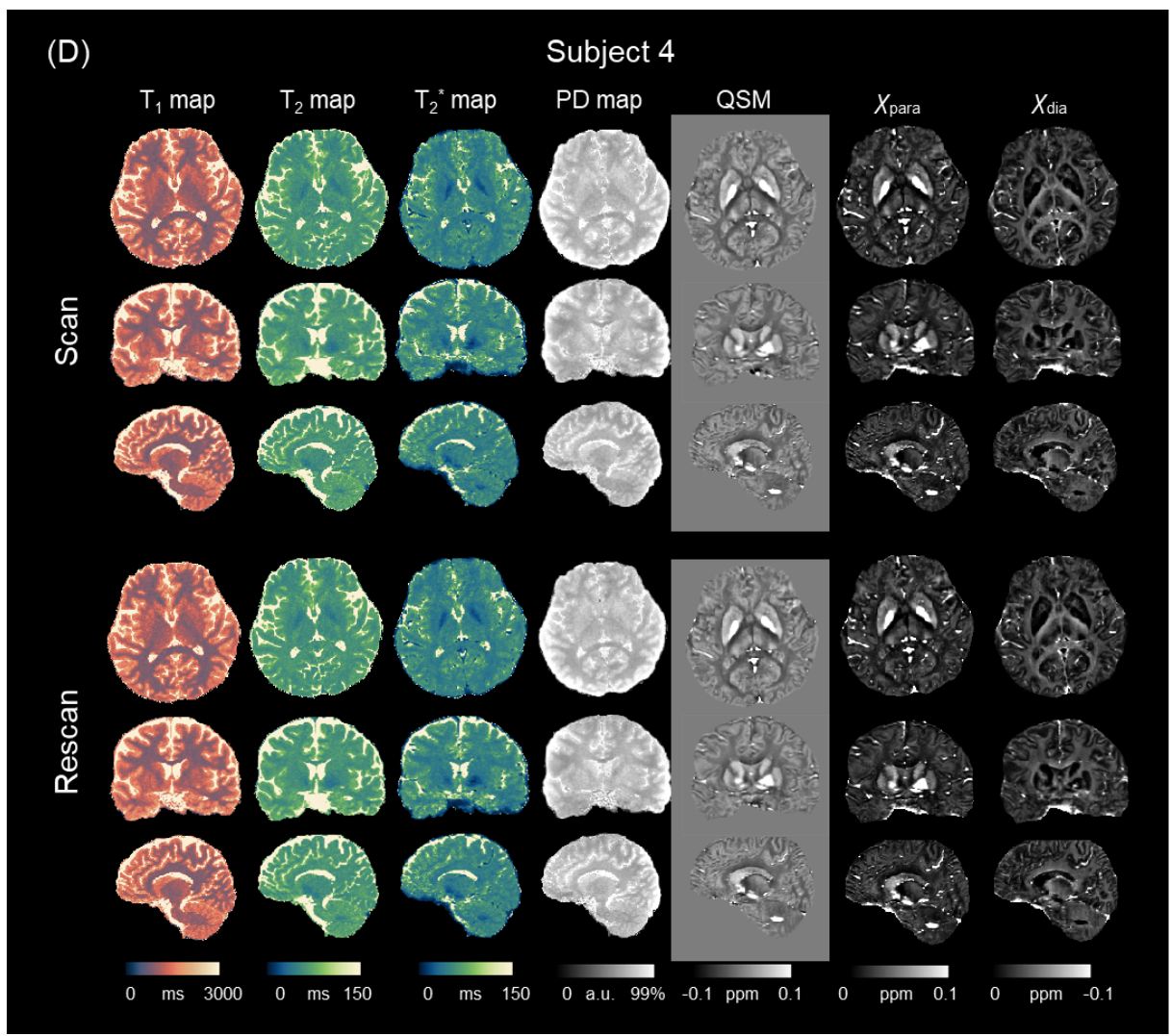

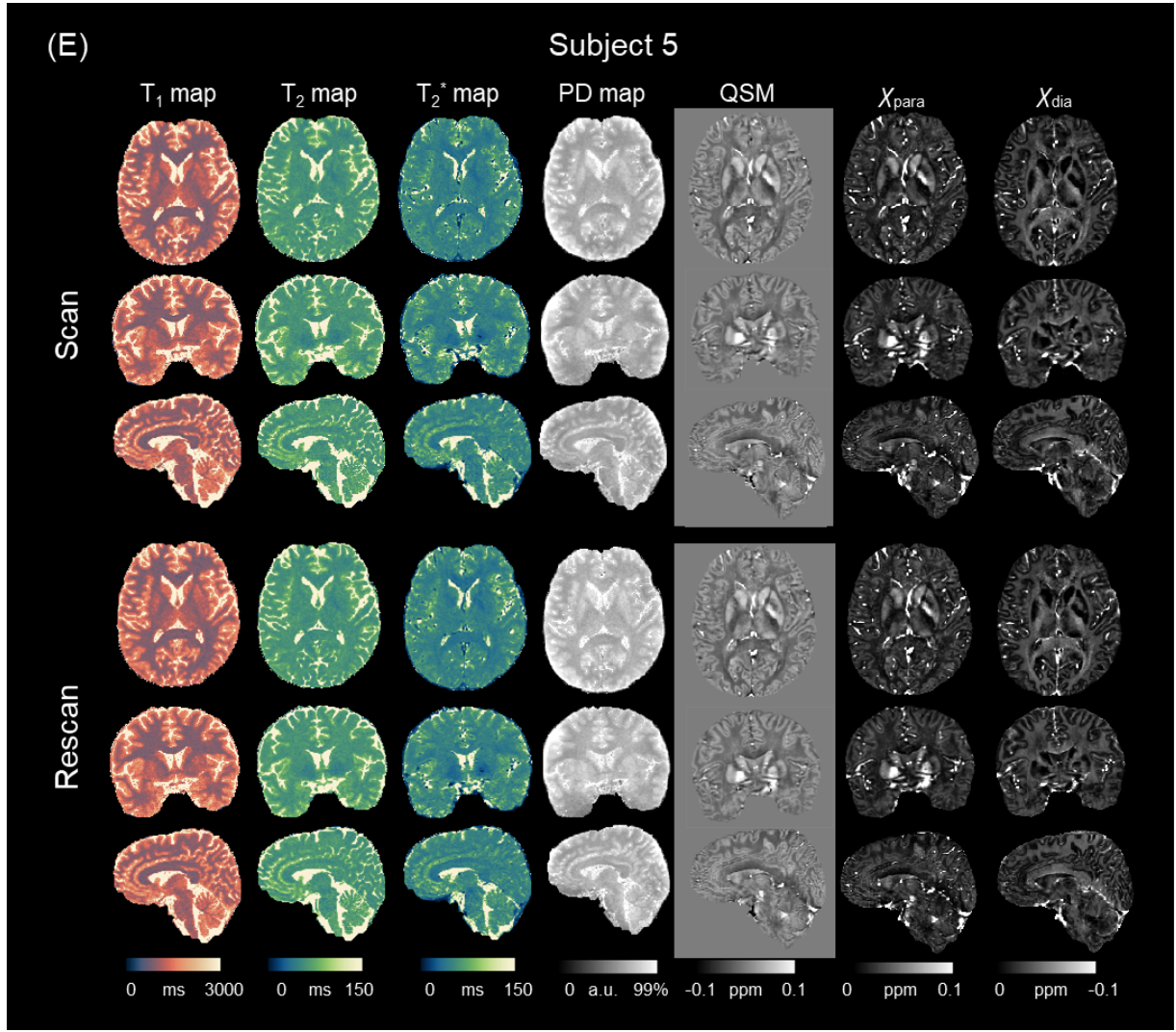

Supporting Information Figure S11. The $T_1$, $T_2$, $T_2^*$, PD, QSM, and susceptibility source separation maps of scan-rescan from five subjects (A-E) using MIMOSA at R=11.8.

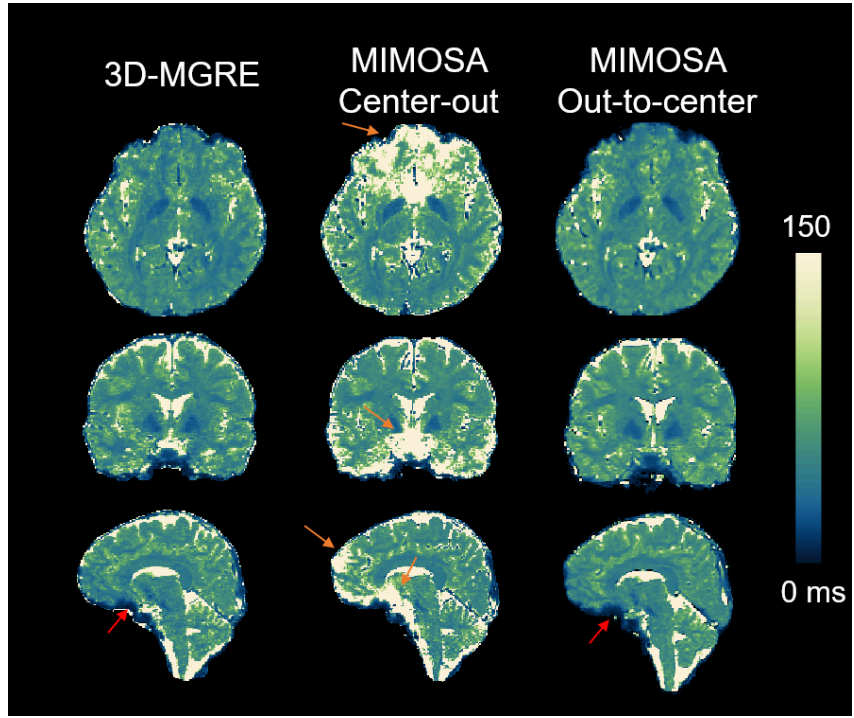

Supporting Information Figure S12. The comparison of $T_2^*$ mapping using 3D-MGRE and MIMOSA with a center-out and out-to-center sampling order of MGRE readout, respectively. 3D-MGRE was performed using the same trajectory of MIMOSA as a reference with the following parameters: FOV = 240×220×196 mm³, matrix size = 192×176×156, resolution = 1.25×1.25×1.26 mm³, TE = [2.7, 7.0, 11.3, 15.6, 19.9, 24.2] ms, TR = 27.5 ms, R = 2.51, TA = 5 min 14 s. The data of MIMOSA was acquired with the following parameters: FOV = 240×220×196 mm³, matrix size = 192×176×156, resolution = 1.25×1.25×1.26 mm³, ETL=127, echo spacing = 5.8 ms, TE = 2.29 ms, flip angle = 4˚, TR of MGRE = 27.5 ms, TEs of MGRE = 2.7, 7.0, 11.3, 15.6, 19.9, 24.2 ms, TR = 6030 ms, R = 2.5, and TA = 9.05 min. 3D-MGRE: 3D multi-echo gradient echo; FOV: Field of view; R: acceleration rate; TA: total acquisition time.